%% file: paper.tex
\PassOptionsToPackage{dvipsnames}{xcolor}
\PassOptionsToPackage{oneside}{geometry}
\documentclass[twocolumn, twocolappendix]{aastex702}
\received{\today}
\shorttitle{Identifying DCOs in \lisa}
\shortauthors{Wagg, Prince, Breivik et al.}
\graphicspath{{figures/}}

\usepackage{lipsum}
\usepackage{physics}
\usepackage{multirow, makecell}
\usepackage{xspace}
\usepackage{natbib}
\usepackage{fontawesome5}
\usepackage{wrapfig}
\usepackage[figuresright]{rotating}
\usepackage{subcaption}
\usepackage{booktabs}
\usepackage{siunitx}
\usepackage{tabularray}

\usepackage[hang,flushmargin]{footmisc}

\usepackage{styles/todo_notes}
\usepackage{styles/abbreviations}

\makeatletter
\renewcommand{\unit}[1]{%
    \,\mathrm{#1}\checknextarg}
\newcommand{\checknextarg}{\@ifnextchar\bgroup{\gobblenextarg}{}}
\newcommand{\gobblenextarg}[1]{\,\mathrm{#1}\@ifnextchar\bgroup{\gobblenextarg}{}}
\makeatother

\DeclareRobustCommand{\msun}{\ensuremath{\unit{M_{\odot}}}\xspace}
\DeclareRobustCommand{\rsun}{\ensuremath{\unit{R_{\odot}}}\xspace}
\DeclareRobustCommand{\mc}{\ensuremath{\mathcal{M}_{\mathrm{c}}}\xspace}

\newenvironment{conclusions}{\begin{enumerate}}{\end{enumerate}}
\newcommand{\conclusion}[2]{\item \textbf{#1}\\#2}

\defcitealias{Wagg+2026:2026arXiv260722814W}{Paper I}

\begin{document}


\title{Charting the Galactic Underworld II: Strategies for Identifying BHs and NSs with \lisa}

\newcommand{\cca}{\affiliation{Center for Computational Astrophysics, Flatiron Institute, 162 Fifth Ave, New York, NY, 10010, USA}}
\newcommand{\cmu}{\affiliation{McWilliams Center for Cosmology and Astrophysics, Department of Physics, Carnegie Mellon University, Pittsburgh, PA 15213, USA}}
\newcommand{\arizona}{\affiliation{University of Arizona, Department of Astronomy \& Steward Observatory, 933 N. Cherry Ave., Tucson, AZ 85721, USA}}
\newcommand{\UW}{\affiliation{Department of Astronomy, University of Washington, Seattle, WA, 98195}}
\newcommand{\Adler}{\affiliation{The Adler Planetarium, 1300 South DuSable Lake Shore Drive, Chicago, IL, 60605, USA}}
\newcommand{\OSU}{\affiliation{Department of Astronomy, The Ohio State University, Columbus, OH 43210, USA}}

\author[0000-0001-6147-5761]{Tom Wagg}
\cca{}
\email{tomjwagg@gmail.com}

\author[0009-0007-0344-3076]{Frank E. Prince}
\email{fep@andrew.cmu.edu}
\cmu

\author[0000-0001-5228-6598]{Katelyn Breivik}
\cmu{}
\email{kbreivik@andrew.cmu.edu}

\author[0009-0009-9373-3097]{Kira Baasch}
\Adler
\email{kbaasch@adlerplanetarium.org}

\author[]{Jorie McDermott}
\OSU
\email{mcdermott.286@buckeyemail.osu.edu}

\author[0000-0002-0147-0835]{Michael Zevin} 
\Adler
\email{mzevin@adlerplanetarium.org}

\correspondingauthor{Tom Wagg}
\email{twagg@flatironinstitute.org, tomjwagg@gmail.com}

\begin{abstract}
    Stellar-mass black holes (BHs) and neutron stars (NSs) observed by space-based gravitational-wave detectors offer constraints on binary stellar evolution that are complementary to those from ground-based detectors. The Laser Interferometer Space Antenna (\lisa) is expected to detect hundreds of double compact objects (DCOs) containing a BH or NS, but distinguishing these sources from the far more numerous double white dwarfs (WDWDs) may prove challenging.
    We use adaptive importance sampling with \cosmic to synthesise large populations of BHBH, BHNS, NSNS, BHWD, and NSWD binaries. We combine these binaries with a Milky Way star-formation history and integrate their Galactic orbits with \cogsworth, repeating our simulations for a grid of variations in uncertain (binary) stellar physics.
    We find that \lisa detections are largely insensitive to the changes in Galactic position resulting from natal kicks. We then assess how well chirp mass, eccentricity, distance from the Galactic plane, and orbital frequency distinguish BH- and NS-hosting DCOs from WDWDs, accounting for \lisa measurement uncertainties. Using chirp mass alone distinguishes only 50--60\% of BHBHs and BHNSs, and ${\sim}40\%$ of NSNSs, but combining this with orbital frequency recovers 90--100\% of BHBHs and BHNSs and around ${\sim}50\%$ of NSNSs, with eccentricity further improving the recovery of NSNSs to around $60$--$70\%$. Distance from the Galactic plane adds little discriminating power, though it can recover a handful of NSWDs that are missed otherwise.
    These methods will help identify a significantly larger number of BHs and NSs in \lisa data that would be missed by a chirp mass cut alone.
\end{abstract}

\keywords{\uat{Gravitational waves}{678}, \uat{Close binary stars}{254}, \uat{Compact objects}{288}, \uat{Stellar evolution}{1599}, \uat{Galaxy dynamics}{591}}




\section{Introduction} \label{sec:intro}

The discovery of GW150914 was soon followed by the recognition that multiband binary black holes (BHBHs) could be discovered as gravitational-wave sources which traverse from the millihertz frequencies observable by space-based detectors to the kilohertz frequencies observable from ground-based detectors \citep{Sesana+2016}. This ignited several studies which sought to understand how space-based detectors like the upcoming Laser Interferometer Space Antenna \citep[\lisa,][]{Amaro-Seoane+2017:2017arXiv170200786A} could provide complementary constraints on the formation environments of BHBHs via measurements of eccentricity \citep{Breivik+2016, Nishizawa+2016}. The focus then expanded to general forecasts for the population of BHBHs discoverable by \lisa \citep{Kremer+2018, Lamberts+2018:2018MNRAS.480.2704L,Kremer+2019:2019PhRvD..99f3003K} and other DCOs containing a BH or NS \citep[e.g.,][]{Lau+2020, Andrews+2020:2020ApJ...892L...9A, COSMIC, Wagg+2022, Tang+2024:2024MNRAS.534.1707T} as well as binaries containing a white dwarf (WD) and a BH or NS \citep[e.g.,][]{Tauris+2018:2018PhRvL.121m1105T, Korol+2024:2024MNRAS.530..844K}. These studies collectively show that while the number of resolved DCO binaries containing a BH or NS will be much lower than the number of resolved WDWDs \citep[e.g.,][and references therein]{Amaro-Seoane+2023:2023LRR....26....2A}, the population is expected to be non-negligible. 

The formation of each NS or BH introduces the potential for natal kicks due to asymmetric mass loss or asymmetries in the explosion mechanism \citep[e.g.,][]{Janka+1994:1994A&A...290..496J,Burrows+1996:1996PhRvL..76..352B}. Each natal kick can alter the properties of the binary at core collapse \cite[e.g.,][]{Flannery+1975:1975A&A....39...61F, Kalogera+1996:1996ApJ...471..352K, Pfahl+2002} as well as modify the systemic motion of the binary on its orbit through the Milky Way \cite[e.g.,][]{Lyne+1994,Wong+2010:2010ApJ...721.1689W,Wagg+2025:2025AJ....170..192W}. Natal kicks thus introduce modifications to both the properties and positions of DCO binaries observable by LISA \citep{McMillan+2026:2026MNRAS.546ag117M}, though the modification of their Galactic populations is unlikely to alter the LISA-resolvable DCO population \citep{Storck+2023:2023MNRAS.521.2368S}. 

Given a population of resolvable DCOs, the problem of \emph{identifying} the DCOs with BHs and NSs as distinct from the WDWDs is thus of considerable importance. \citet{Sesana+2020} showed that chirp mass measurements for BHBHs with observable frequency evolution simulated by \citet{Lamberts+2018:2018MNRAS.480.2704L} can be used to distinguish BHBHs from the lower-mass NSNS and WDWD binaries. They also showed that while higher-frequency BHBHs are more likely to have three-dimensionally-localised populations, lower-frequency BHBH binaries can be distinguished from lower-chirp-mass binaries through electromagnetic (EM) follow-up studies for nearby systems containing WDs. 

More recently, building on the work of \citet{Seto+2001:2001PhRvL..87y1101S}, \citet{Moore+2024:2024MNRAS.531.2817M} showed that eccentricity measurements via multiharmonic heterodyning analysis of the GW frequency harmonics produced by eccentricity can aid in the discovery of NSNS binaries as well as potentially resolve each NS's mass through periapse precession measurements. \citet{Middleton+2026:2026ApJ..1010...19M} applied a Bayesian analysis pipeline to the simulations from \citet{Wagg+2022} to show that individual component masses can be measured in $10$--$50\%$ of the resolved DCOs using the periapse measurements. For NSWD binaries, \citet{Korol+2026:2026A&A...705A.154K} showed that while these systems are difficult to distinguish from WDWDs with GW data alone, their EM signatures may be especially useful. Finally, \citet{Tay+2026:2026arXiv260306341T} applied a machine learning classifier to DCOs, neglecting GW measurement uncertainties, to show that binaries containing NSs and BHs only are easy to distinguish from WDWDs with mass measurements while NSWDs require optimizations to the classifier. 

This paper is the second in a series exploring the Milky Way's population of compact objects, self-consistently accounting for binary evolution and Galactic orbit integration. The first paper \citep[hereafter Paper~I]{Wagg+2026:2026arXiv260722814W} simulated the intrinsic population of Milky Way BHs, examining their kinematics, rates, and demographics. \citetalias{Wagg+2026:2026arXiv260722814W} found that BHBHs are the most common BH binary in the Galaxy and assessed their detectability with microlensing. However, only a small fraction of these binaries are compact enough to emit detectable gravitational waves in the mHz band. Here we instead focus on \lisa-detectable DCOs, extending our simulations to systems containing NSs and to binaries in which a BH or NS is paired with a WD.

In this work, we jointly explore how binary mass exchange and natal kicks modify the properties and spatial distribution of mHz GW sources and investigate how these effects impact the ability to distinguish DCOs from the thousands of WDWD binaries LISA will discover. In Section~\ref{sec:sims}, we describe our simulation setup and model variations. In Section~\ref{sec:kicks} we describe the impact of accounting for natal kicks on the Galactic orbit evolution of each binary, and thus the present-day spatial distribution of our simulated DCO populations. We then explore different strategies for identifying DCOs in the LISA data stream in Section~\ref{sec:dist_methods} and discuss other relevant work and caveats in Section~\ref{sec:discussion}. We finish with our conclusions in Section~\ref{sec:conclusions}. 

\section{Creating a \lisa-detectable population of Milky Way DCOs}
\label{sec:sims}
In this section we explain our methods for creating a population of Milky Way DCOs that are detectable with \lisa. Our methodology builds upon \citealt{Wagg+2022}, using a similar star-formation history and population synthesis code, while additionally self-consistently accounting for the kinematics of each DCO with the \cogsworth package \citep[][v4.0.2]{cogsworth_apjs}. \cogsworth combines rapid binary population synthesis using \cosmic \citep[][v4.2.1]{COSMIC} with the galactic dynamics using \gala \citep[v1.11.0]{Gala}.

We first describe our implementation of the \stroop algorithm \citep{STROOPWAFEL} in \cosmic and the setup we use for generating a population of hundreds of millions of binaries across metallicities in Section~\ref{sec:stroop_explained}. Next, in Section~\ref{sec:connect-to-mw}, we detail how we connect this population to the Milky Way, using \cogsworth to apply a model for the Galaxy's star-formation history and gravitational potential. Finally, we explain how we select the \lisa-detectable subset of binaries with \legwork \citep[][v1.0.0]{LEGWORK_joss, LEGWORK} in Section~\ref{sec:lisa-detection-methods}. All code and simulations are openly available on Zenodo (see Section~\ref{sec:availability}).

\subsection{Synthesising rare DCOs with adaptive importance sampling} \label{sec:stroop_explained}

Given the rarity of DCO formation, regular Monte Carlo sampling may not produce a representative population of objects without an extremely large sample. We therefore implemented the \stroop adaptive importance sampling algorithm \citep{STROOPWAFEL} in version 4.2 of \cosmic. This implementation can be applied more broadly to any rare population, and adaptively sample based on initial masses, separation, eccentricity, metallicity, and natal kicks. It is documented in a series of tutorials that can be found in the \href{https://cosmic-popsynth.github.io/COSMIC/}{\cosmic documentation}.\footnote{\url{https://cosmic-popsynth.github.io/COSMIC/}}

\subsubsection{Algorithm description}

The \stroop algorithm operates in three phases and can be summarised as follows:
\begin{enumerate}
    \item \textit{Exploration} -- Random samples are drawn from the prior distributions, evolved with \cosmic, and if the system forms an object of interest it is marked as a hit.
    \item \textit{Adaptation} -- A multivariate Gaussian component is placed at each hit location in parameter space.\footnote{The width of each Gaussian is derived from the local density of the prior so that the proposal is appropriately broad regardless of the parameter’s scale.} These components combine to form a mixture model that maps out the formation space of hits.
    \item \textit{Refinement} -- After the adaptation, new binaries are drawn from the mixture model instead of the prior. Each simulated system receives an importance weight, $w(x)$, which is defined as
\end{enumerate}
\begin{equation}
    w(x) \equiv \frac{\pi(x)}{Q(x)}, \qquad Q(x) = f_e \pi(x) + (1 - f_e) q(x),
    \label{eq:ais_weights}
\end{equation}
where $\pi(x)$ is the prior probability density, $q(x)$ is the Gaussian mixture density, and $f_e$ is the fraction of systems drawn during exploration. Using these weights, any statistic computed on the hit population is an unbiased estimator of the corresponding prior-weighted quantity. For more details on this algorithm we refer the reader to \citet{STROOPWAFEL}, and additions in \citet{floris_thesis} and \citet{loki_thesis}.

\subsubsection{\stroop setup for DCOs}

We sample a population of DCOs in 50 bins of metallicity, $Z$, log-spaced with $Z \in [10^{-4}, 0.03]$ where we assume that $Z_{\odot} = 0.014$ \citep{Asplund+2009:2009ARA&A..47..481A}. For each bin we assume the following priors for \stroop sampling. We draw masses of the primary (initially more massive) star in each binary, $m_1$, following the broken power law initial mass function (IMF) from \citet{Kroupa+2001:2001MNRAS.322..231K}, such that $p(m_1) \propto m_1^{\alpha_{\rm IMF}}$ and sampling only stars with $m_1 \in [m_{1, \rm min}, 150] \msun$. The high-mass slope (for stars with $m_1 > 1 \msun$) of this IMF has $\alpha_{\rm IMF} = -2.3$. We set $m_{1, \rm min}$ differently for each DCO type, using $4\msun$ for NSWDs, $5\msun$ for NSNSs, $15\msun$ for BHWDs, $16\msun$ for BHNSs, and $19\msun$ for BHBHs. We sample mass ratios, $q \equiv m_2 / m_1$, uniformly in $[q_{\rm min}, 1]$, where $q_{\rm min}$ is set such that the minimum secondary mass is $0.08 \msun$ \citep{Mazeh+1992:1992ApJ...401..265M, Goldberg+1994:1994AA...282..801G}. Eccentricities, $e$, are sampled following \citet{Sana+2012:2012Sci...337..444S} such that $p(e) \propto e^{-0.45}$ with $e \in [0, 0.9]$, where the upper limit is chosen to avoid Roche lobe overflow at pericentre (at birth). For orbital periods, $P$, we follow the distribution of \citet{deMink+2015:2015ApJ...814...58D}, which is an extrapolation of \citet{Sana+2012:2012Sci...337..444S} to lower masses, such that $p(P) \propto (\log_{10}(P / \unit{days}))^{-0.55}$ with $\log_{10}(P / \unit{days}) \in [0.15, 5.5]$. We assume the binary fraction scales with primary mass following \citet{Offner+2023:2023ASPC..534..275O}.

We simulate each DCO type separately, so that the adaptive importance sampling can most efficiently target the progenitors of each DCO type. We sample a population of 2 million binaries in each metallicity bin, giving a total of 100 million binaries for each DCO type.


\subsubsection{\cosmic population synthesis assumptions}\label{sec:cosmic_explained}

\begin{table*}
    \centering
    \begin{tblr}{colspec={lX}, cells={valign=m}, hline{2}={solid}}
        \hline
        Label & Description \\
        Fiducial & - \\
        C14 $q_{\rm crit}$ & Follow for \citet{Claeys+2014:2014AA...563A..83C} for determining the critical mass ratio for mass transfer stability \\
        Optimistic CE & Allow binaries to survive a common envelope initiated by donor without a distinct core-envelope boundary \\
        $\alpha_{\rm CE} = 0.1$ & Decrease the envelope ejection efficiency of common-envelope events by $10 \times$ \\
        $\alpha_{\rm CE} = 10$ & Increase the envelope ejection efficiency of common-envelope events by $10 \times$ \\
        M25 remnants & Follow \citet{Maltsev+2025:2025AA...700A..20M} and \citet{Willcox+2025:2025arXiv251007573W} for converting pre-supernova carbon-oxygen cores to remnant masses \\
        No ECSN & Turn off electron-capture supernovae (which, by default in the fiducial model, produce distinct NS masses and have weaker natal kicks) \\
        \hline
    \end{tblr}
    \caption{A description of the model variations simulated in this work. See Section~\ref{sec:cosmic_explained} for a full description of the fiducial settings and more details on each variation.}
    \label{tab:models}
\end{table*}

To perform binary evolution simulations using \cosmic, we must define how \cosmic treats assumptions for wind mass loss, CO formation and natal kicks, and binary interactions via tides and Roche-lobe-overflow mass transfer. We first consider a fiducial model based on the default settings for \cosmic\footnote{These settings are maintained in the  \href{https://github.com/COSMIC-PopSynth/COSMIC/blob/develop/examples/Params.ini}{COSMIC GitHub repository}.} from which we explore variations that impact the number and properties of DCO binaries that LISA may observe. While we do not discuss every setting in detail here for brevity, we highlight the key assumptions that either imprint heavily on the DCO population, or have recently changed as default \cosmic assumptions. Our full list of settings is available on Zenodo (see Section~\ref{sec:availability}).

We use the widely applied model set for metallicity-dependent line-driven winds described in \citet{Vink+2001} for hydrogen-rich OB stars and \citet{Vink+2005} for Wolf-Rayet stars. However, for stars which have luminosities and temperatures beyond the Humphreys-Davidson limit, we apply the winds described in \citet{Hurley+2002} for luminous-blue-variable-like winds rather than the fixed rate of $1.5\times10^{-4} M_{\odot}/\rm{yr}$ that was used up to v3.7.2 in \cosmic. This leads to higher wind mass loss rates for stars above $\sim35\,M_{\odot}$ at solar metallicity. 

We assign a mass transfer stability model by applying critical mass ratios, $q_{\rm{crit}}=M_{\rm{donor}}/M_{\rm{accretor}}$, which define the boundary above which mass transfer becomes dynamically unstable as a function of the donor star's evolutionary stage. Our fiducial model follows the set of critical mass ratios set by $\texttt{qcflag}=5$ in the \texttt{COSMIC} settings. Consistent with \citet{Neijssel+2019}, we convert the $\zeta \equiv \pdv{\ln R}{\ln M}$ values to $q_{\rm{crit}}$ under the assumption of conservative mass transfer where donors on the Main Sequence and Hertzsprung gap have $\zeta_{\rm MS} = 2.0$, $\zeta_{\rm HG} = 6.5$ respectively, corresponding to $q_{\rm{crit,MS}}=1.717$ and $q_{\rm{crit,HG}}=3.825$. For stars on the giant and asymptotic giant branches, we use the parametric functions for $q_{\rm{crit,GB/AGB}}$ following \citet{Hjellming+1987:1987ApJ...318..794H}. Finally, for core helium burning stars, we assume that $q_{\rm{crit,CHeB}}=3.0$ following \citet{Hjellming+1989:1989PhDT.........7H}, and for any stripped helium stars, we assume that all mass transfer is stable such that $q_{\rm{crit,He}}=\infty$. 

For binaries that do undergo common-envelope evolution, we apply the $\alpha$-$\lambda$ prescription \citep{Webbink+1984:1984ApJ...277..355W, deKool+1990:1990ApJ...358..189D}. $\alpha_{\rm CE}$ is the fraction of orbital energy that is used to unbind the envelope, which we set to 1 in the fiducial model. $\lambda$ is the envelope binding energy parameter and is fit to the stellar models of \citet{pols:98}, as described in Appendix A of \citet{Claeys+2014:2014AA...563A..83C}.
By default, we assume that donors that are crossing the Hertzsprung gap have not yet built up a distinct core--envelope boundary and thus do not successfully eject the shared envelope regardless of the binary properties at the onset of RLOF. This is often referred to as the ``pessimistic'' common envelope treatment \citep[e.g.,][]{Dominik+2012}. 

We assume that compact objects form according to the \citet{Fryer+2012:2012ApJ...749...91F} ``delayed'' prescription which produces COs with a continuous mass function, and we assign the maximum NS mass (and thus minimum BH mass) to be $M_{\rm{NS,max}}=3\,M_{\odot}$. The natal kick for each CO formed through a core-collapse event is drawn randomly from a log-normal distribution in $\ln$($v_{\rm{kick}}/\rm{km\,s^{-1}}$) with $\mu=5.60$ and $\sigma=0.68$ following \citet{Disberg+2025:2025ApJ...989L...8D}, then decreased by the fraction of the stellar envelope that falls back onto the proto-CO as in \citet{Fryer+2012:2012ApJ...749...91F}. This choice leads to stronger kicks with decreasing CO mass, and no asymmetric natal kick for BHs with masses above $\sim14\,M_\odot$. For COs that are formed through an electron-capture supernova (ECSN), or ultra-stripped supernova (USSN), we assume that natal kicks are weaker due to less asymmetry in the explosion. For these supernovae, we draw natal kicks from a Maxwellian with a scale parameter of $20\unit{km}{s^{-1}}$ \citep[e.g.,][]{Igoshev+2020}.


\paragraph{Variations}
We explore variations from the fiducial model above in the treatment of mass transfer stability, common-envelope evolution, remnant masses, and ECSNe. Our variations are summarised in Table~\ref{tab:models}. 

The stability of Roche-lobe overflow (RLOF) mass transfer plays a key role in the amount of mass exchange versus mass loss in a binary, thus directly influencing whether each stellar component has enough mass at core collapse to produce a CO or not. We explore a variation in assumptions for the critical mass ratio at which mass transfer becomes dynamically unstable and initiates a common envelope evolution denoted as $\texttt{qcflag=2}$ in \cosmic. In this model, consistent with \citet{Claeys+2014:2014AA...563A..83C}, main-sequence donors have a reduced critical mass ratio of $q_c=1.6$, while Hertzsprung-Gap donors have an increased critical mass ratio of $q_c=4.0$. The critical mass ratio of giant and asymptotic giant branch stars is assigned following the mass-dependent model described in \citet{Claeys+2014:2014AA...563A..83C}. Finally, stripped helium main sequence stars and stripped helium Hertzsprung gap stars have $q_c=3$ and $q_c=4$ respectively. 

Common-envelope evolution is an extremely difficult process to model faithfully \citep[e.g.][]{Ivanova+2013}. To explore the effects of our treatment of the common envelope, we consider three variations. In the first, we use the ``optimistic'' common-envelope treatment, in which we do \textit{not} force a merger for common envelopes initiated by a Hertzsprung gap donor stars. In the other two variations, which apply the ``pessimistic'' common envelope treatment (as in the fiducial model), we set the envelope ejection efficiency to a very low ($\alpha_{\rm{CE}}$ = 0.1) and very high ($\alpha_{\rm{CE}}=10)$ value respectively. A lower efficiency means that binaries must shrink more before the envelope can be unbound, which leads to more stellar mergers, but also tighter binaries for those that successfully eject their envelope.

We finally consider variations on assumptions for CO formation. First, we consider a different remnant mass prescription inspired by the ``core-compactness peaks'' introduced in \citet{Schneider+2023:2023ApJ...950L...9S} that follows \citet{Maltsev+2025:2025AA...700A..20M} and its implementation in population synthesis as described by \citet{Willcox+2025:2025arXiv251007573W}. Specifically we extrapolate the fits of \citet{Maltsev+2025:2025AA...700A..20M} across our full metallicity range, assume a fallback fraction of $f_{\rm fallback} = 0.5$ where $M_{\rm{rem}} = 1.4\,M_{\odot} + f_{\rm{fallback}}(M_{\rm core}- 1.4 M_{\odot})$, and assume that $10\%$ of remnants in the partial fallback regime form BHs. We consider an additional variation where we remove the possibility for ECSNe, which form from stars with enclosed helium masses between $1.6$--$2.25\,M_{\odot}$ following \citet{Belczynski+2008}. Traditionally, ECSNe are overrepresented in the DCO population containing NSs since they are associated with low kicks that do not unbind the binary. In this variation, stripped helium star supernovae are the only pathway for producing NSs via explosions that impart small natal kicks to the newly formed CO. 

\subsection{Connecting synthetic populations to the Milky Way's star-formation history}\label{sec:connect-to-mw}

The intrinsic population of merging DCOs in the Milky Way can be simulated by drawing from our synthetic populations. The synthetic population of DCOs is sampled log-uniformly in metallicity and each binary is evolved for 12 Gyr, our assumed age of the Milky Way. In reality, the Milky Way has a more complex metallicity distribution, which is tied to the lookback times of binaries and their positions. We use a model for the Milky Way's star-formation history (SFH) from \citet{Sanders+2015:2015MNRAS.449.3479S}. This model includes both a thin and thick disc, which each have different kinematics and lookback time distributions \citep[see][Eq.~3, 10, and 11]{Sanders+2015:2015MNRAS.449.3479S}. Critically, this SFH uses an action-based distribution function that ensures the velocities sampled from the distribution are consistent with positions and the Galactic potential. This self-consistency ensures that any difference in kinematics at present day is due to the influence of supernova (SN) kicks, rather than sources artificially sinking into the potential. We link the positions and birth times drawn from this model to metallicities following \citet{Frankel+2018}, which provides a metallicity-radius-time relation that accounts for radial migration and the inside-out growth of the Galaxy.

We create a realisation of the Milky Way by evolving a sample of the synthetic populations through the Galaxy. We draw $N_g = $ 500,000 binaries randomly from the synthetic populations, weighted by $p_Z$, which is the probability of forming a star with a metallicity $Z$ in the assumed SFH. For each of these binaries we sample from the SFH and assign the binary an initial Galactic position, velocity, and birth time that correspond to its metallicity.

\subsection{Attaining a present-day intrinsic DCO population and selecting \lisa detectable binaries}\label{sec:lisa-detection-methods}

We attain an intrinsic population of DCOs by evolving binaries from their formation states until present-day. For BHBHs, BHNSs, and NSNSs, we evolve the binary orbits of DCOs due to the emission of gravitational-waves until present day using \legwork \citep{LEGWORK_joss,LEGWORK}.\footnote{For efficiency, we mask out any of the $N_g$ DCOs that would not yet have formed (for those very recently born) or have already merged (for those born early in the Galaxy's history and/or with short merger times).} This evolution follows the equations in \citet{Peters+1963} and \citet{Peters+1964}, which outline how gravitational waves will shrink and circularise a binary using the post-Newtonian approximation.

For NSWDs and BHWDs, we additionally account for the potential \textit{widening} of binary orbits as a result of Roche lobe overflow (RLOF). These systems tighten due to GWs until the WD overflows its Roche lobe, initiating mass transfer onto the NS or BH. Due to the mass ratio of the system, this mass transfer counteracts the tightening from gravitational-wave emission. Over time the systems gradually widen, eventually reducing the WDs to planet-like remnants orbiting a BH or millisecond pulsar. We refer the reader to \citet{Tauris+2018:2018PhRvL.121m1105T} for more details on mass-transferring NSWDs in the \lisa band. For our simulations, we use \cosmic to calculate present-day orbital configuration and masses for any NSWDs or BHWDs that experience RLOF before present day. This subpopulation accounts for around 60\% of the detectable NSWDs and 20\% of the detectable BHWDs in our simulations.

Next, using \cogsworth \citep{cogsworth_apjs}, we evolve the \textit{Galactic} orbits of each DCO through the Galactic potential, using version 2 of the Milky Way potential from \gala \citep{Gala}. \cogsworth accounts for the impact of each SN kick (whether as a result of symmetric mass loss from the collapsing star, or from a natal kick due to asymmetry in the ejecta) on the Galactic orbit of a binary. Finally, given the present-day parameters of the system (primary mass $m_1$, secondary mass $m_2$, separation $a$, eccentricity $e$, and distance $D$), we use \legwork to compute the position, orientation, polarisation-averaged \snr for each source. The full derivation of this \snr calculation is detailed in \citet{LEGWORK}. We assume that DCOs with an \snr above 12 are individually resolvable by \lisa. This threshold is more stringent than previous works investigating \lisa DCOs \citep[e.g.,][]{Korol+2020, COSMIC, Wagg+2022}, but has been shown to be more effective during the \lisa data challenge \citep{Bandopadhyay+2025:2025PhRvD.111d4039B}.

We calculate the \snr under several assumptions to make our results useful for several studies. We calculate \snr values for \lisa, varying the mission observation times from 4 to 8 years. We also calculate the \snr using both the formation locations of DCOs \citep[as has been done in several previous works, e.g.,][]{Nelemans+2001, Wagg+2022} and the more accurate present-day positions after Galactic orbit integration. For each of these assumptions we compute the detectable fraction of merging DCOs that form in the Milky Way instance, $f_{\rm detect}$ as
\begin{equation}
    f_{\rm detect} = \frac{\sum_i^{N_g} w_i \, \theta_{{\rm detect}, i}}{\sum_i^{N_g} w_i},
\end{equation}
where $w_i$ is the adaptive importance sampling weight of each DCO (see Eq.~\ref{eq:ais_weights}), and $\theta_{{\rm detect}, i} = 1$ if the DCO is detectable and zero otherwise.

We repeat this calculation for 2560 realisations of the Milky Way (for each DCO type and physics variation) to obtain a large population of detectable DCOs and estimate the uncertainty on this population.

\subsection{Normalising to the Milky Way population}

Each Milky Way instance that we evolve represents a fixed population of $N_g$ binaries that must be normalised to obtain a true Milky Way detection rate. We follow the methodology outlined in Appendix B of \citet{Wagg+2022} to compute the number of DCOs merging in a Hubble time that have formed in the Milky Way for each DCO type ($N_{\rm target, MW}$). We assume that the total star-forming mass of the Milky Way is $M_{\rm SF, MW} = 10.4 \times 10^{10} \msun$ \citep[][Appendix B.2]{Wagg+2022}. We show the calculated values of $N_{\rm target, MW}$ for each DCO type and model variation in Table~\ref{tab:n_targets}.
\begin{table}
    \centering
    \input{tables/n_targets}
    \caption{The number of DCOs that merge in a Hubble time that have formed over the history of the Milky Way, for each model variation.}
    \label{tab:n_targets}
\end{table}
Finally, we compute
\begin{equation}
    N_{\rm detect} = f_{\rm detect} \cdot N_{\rm target, MW},
\end{equation}
which is the number of detectable sources in the Milky Way.

\section{The impact of SN kicks on \lisa-detectable DCOs}
\label{sec:kicks}

In this Section we examine the impact of SN kicks on the population of \lisa-detectable DCOs. We first show how SN kicks kinematically heat the Galactic orbits of DCOs, increasing the scale height of the population, particularly for systems that do not contain a WD (\S\ref{sec:kicks-effect-on-positions}). However, we then highlight that, despite this kinematic heating, \lisa detections are mostly insensitive to the change in position (\S\ref{sec:insensitivity-to-position}).

\subsection{SN kicks kinematically heat \lisa-detectable DCOs}\label{sec:kicks-effect-on-positions}

\begin{figure}
    \centering
    \includegraphics[width=0.85\columnwidth]{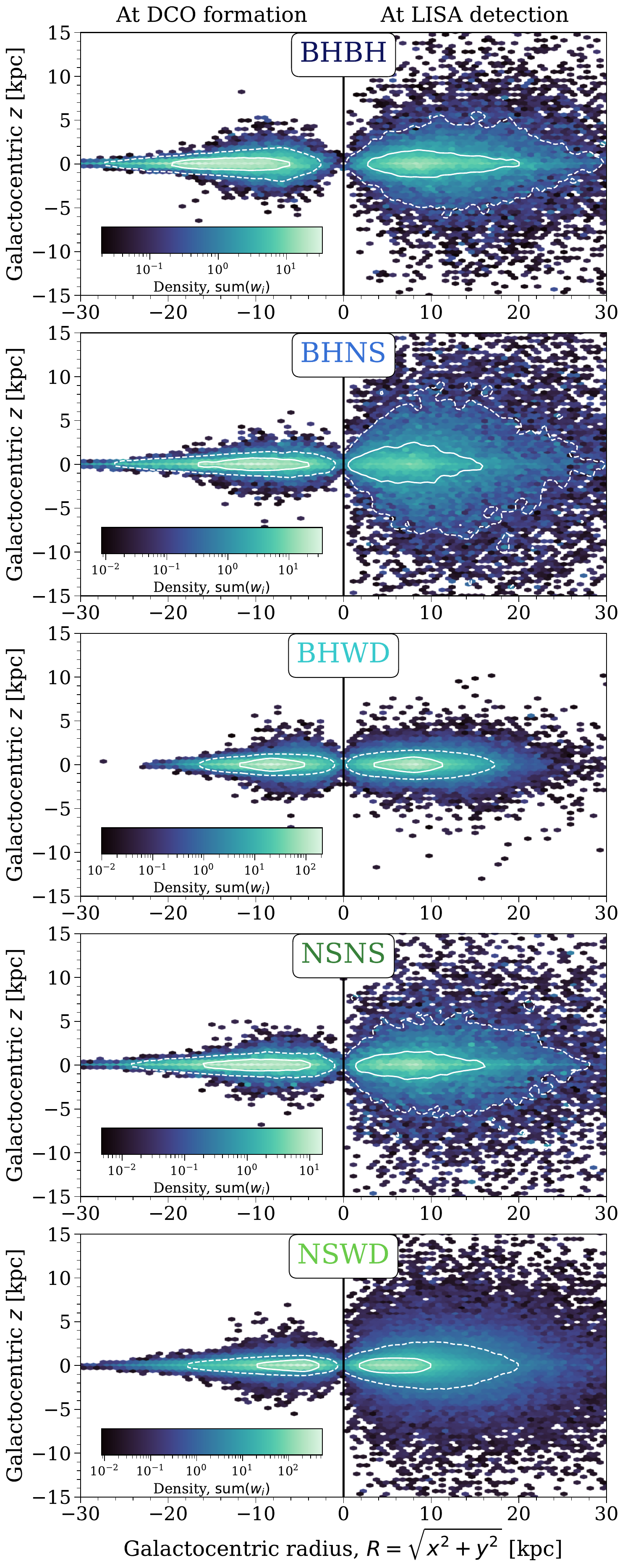}
    \caption{SN kicks broaden the present-day position on DCOs that contain a BH or NS. Each panel shows 2D density distributions (weighted by \stroop weights) for the population of DCOs detectable by a 8-year \lisa observation duration. The left sides show the location at the moment of DCO formation. The right sides show the position of DCOs at present day, after integrating their Galactic orbits through the Milky Way potential. The solid white lines show the 50\% contour and the dashed lines show the 90\% contour for each population. Note that the colour scale is different for each panel.}
    \label{fig:dco_positions}
\end{figure}

The distribution of \lisa-detectable DCOs across the Milky Way is significantly broader after accounting for the impact of SN kicks. In Figure~\ref{fig:dco_positions} we show the positions of \lisa-detectable DCOs, both at the moment of binary formation (left panels), and at present-day, after integrating their Galactic orbits through the Milky Way potential (right panels). For each DCO type, the formation locations approximately follow the overall star-formation history. The slight variations between panels for the formation location are because different DCO types form from different metallicities, which are correlated with position in the star-formation history model. The right panels show that the velocities imparted by SN kicks considerably increase the scale height of DCOs containing a BH or NS \citep[in agreement with][Figure 2]{McMillan+2026:2026MNRAS.546ag117M}.

The impact of SN kicks on DCO position is weaker for DCOs containing a WD. The scale height of BHBHs increases by $3.0\times$ when accounting for SN kicks; similarly, the scale height for BHNSs increases significantly by $6.9\times$, and NSNSs by $4.2\times$. The increase for NSNSs is less than for BHNSs because (a) their binding energy is lower, biasing towards weaker kicks and (b) NSs in NSNS binaries are more often formed through ECSNe, which we assume receive weaker natal kicks (see Section~\ref{sec:cosmic_explained}). However, by comparison, the NSWD and BHWD scale heights only increase by $1.6\times$ and $2.7\times$ respectively. The weaker effect for the WD-containing DCOs is because they cannot survive strong SN kicks, as we explain below.

\begin{figure}
    \centering
    \includegraphics[width=\columnwidth]{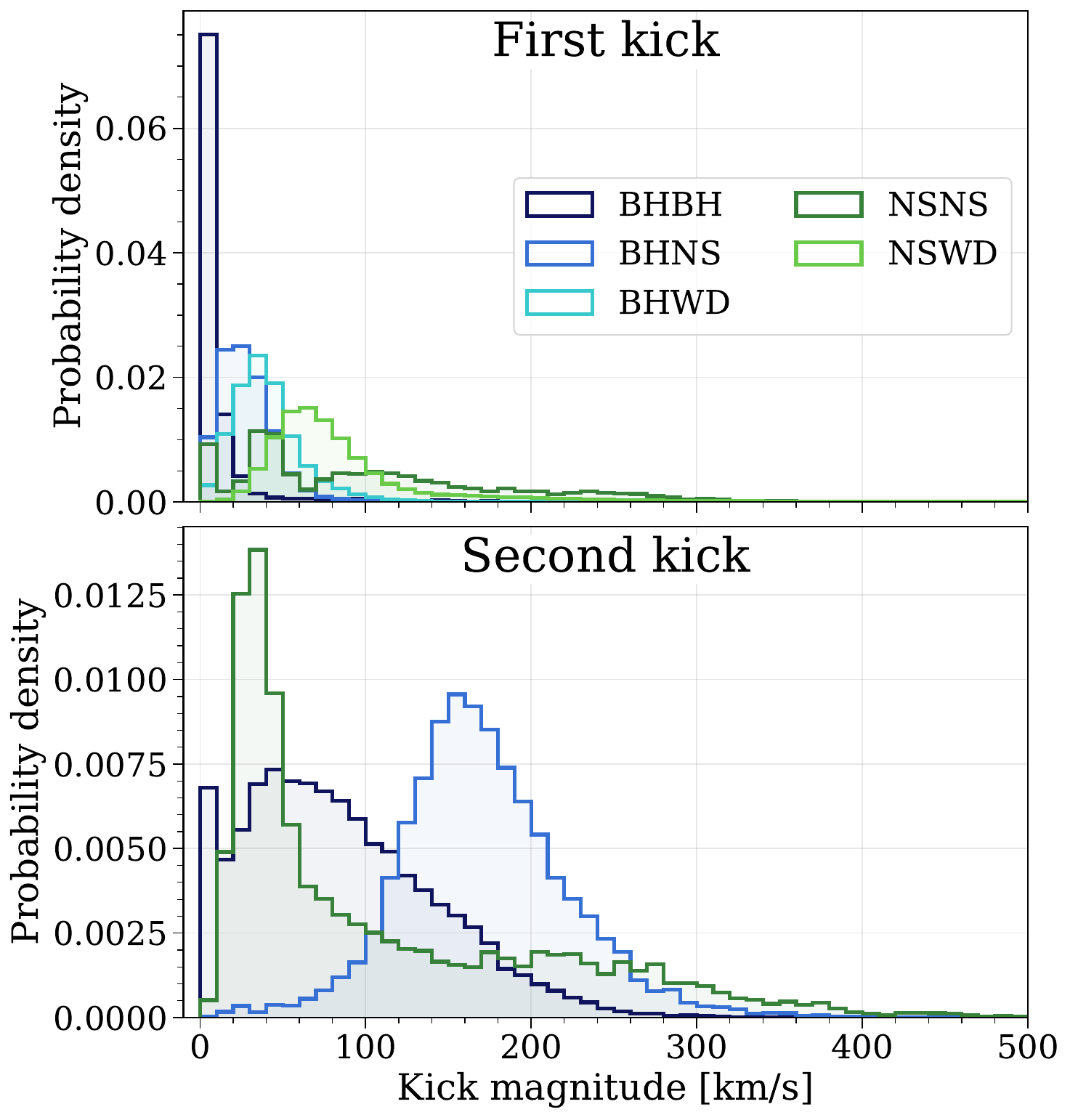}
    \caption{The second SN kick magnitude typically dominates the first for \lisa-detectable DCOs. Each panel shows normalised histograms for the kick magnitude (combining both Blaauw and natal kicks) applied to each DCO. The top panel shows the magnitudes for the first kicks, while the bottom panel shows the second kick (hence does not show BHWDs and NSWDs). Note that the $y$-axis limits change between panels.}
    \label{fig:kick_mags}
\end{figure}

The second SN kick in a bound DCO progenitor is typically much larger than the first \citep[see e.g.][]{Vigna-Gomez+2018}. In Figure~\ref{fig:kick_mags}, we show the distribution of SN kick magnitudes (combining both Blaauw and natal kicks) for the \lisa-detectable population of each DCO type. The top panel shows the kick from the first SN, while the bottom panel shows the second SN kick (and hence NSWDs and BHWDs are not shown in this panel). This figure shows that there is a stark contrast in the kick magnitude applied by the first and second SNe; the second kick is typically much stronger for bound DCOs. 

The second SN kick being stronger is primarily a \textit{selection} effect. For a DCO to be detectable by \lisa, it needs to not only remain bound, but also remain tight enough to merge in a Hubble time. A large SN kick works contrary to both of these criteria \citep[e.g.,][]{Vigna-Gomez+2025:2025A&A...701L...3V}. As a result, in general most binaries are disrupted by the first supernova in the system \citep[e.g.][]{Renzo+2019:2019A&A...624A..66R, Wagg+2025:2025AJ....170..192W}. For this reason, SN natal kicks of \lisa-detectable DCOs are smaller than what one would expect from a random draw from our assumed natal kick distribution (which has a median kick of ${\sim}270\unit{km}{s^{-1}}$). However, this selection is moderately relaxed for the second SN, which occurs after the system has typically experienced mass transfer which shrinks the orbit and increases the binding energy. For example, for \lisa-detectable BHBHs, the median separation at the moment of the first SN in the fiducial model is ${\sim}900\rsun$, but reduces to ${\sim}3.5\rsun$ by the second SN (93\% of systems have experienced a common-envelope by this point). In addition, Blaauw kicks are stronger for tighter systems, so even completely symmetric secondary explosions can often receive stronger kicks. As a result of these effects, for \lisa-detectable BHBHs, the median kick magnitude increases from ${\sim}1\unit{km}{s^{-1}}$ in the first SN to ${\sim}60\unit{km}{s^{-1}}$ in the second.

\begin{figure*}
    \centering
    \includegraphics[width=\textwidth]{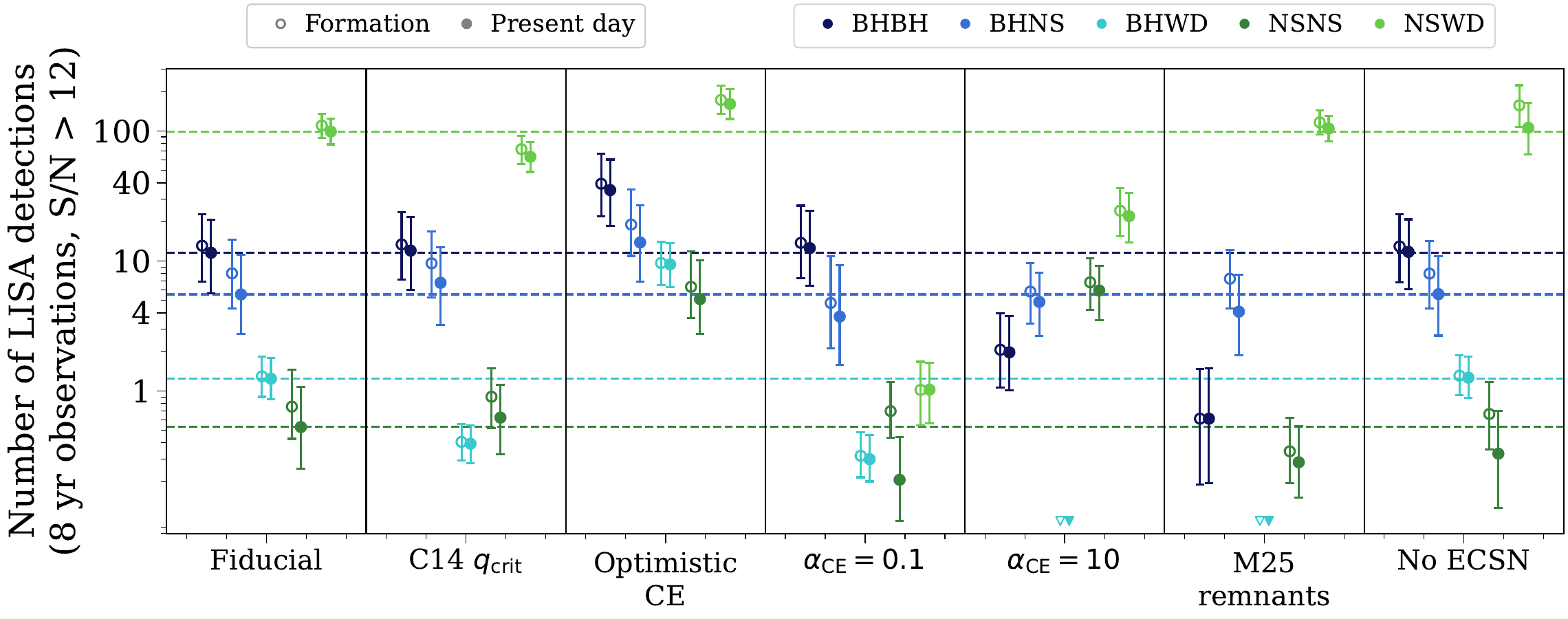}
    \caption{The \lisa detection rate of DCOs is mostly insensitive to the change in position due to SN kicks. Each scatter point shows the median number of detectable DCOs for a 8-year \lisa mission, for each DCO type and model variation. Open points show the number of detections when using the formation location in the \snr calculation, while the closed points show the number of detections when using the present-day position. The error bars show the 5th and 95th percentiles from 2560 Milky Way instances. Each group of points corresponds to a model variation. The exact values are given in Tables~\ref{tab:detections_init}~\&~\ref{tab:detections_final}.}
    \label{fig:n_detect}
\end{figure*}

NSWDs and BHWDs never experience a second SN, which means that they are more kinematically similar to the stellar distribution. The median separation of \lisa-detectable BHBH, BHNS, and NSNSs at the moment of their second SN is $\sim 3.5 \rsun,\ 8.1 \rsun,\ 16.6\rsun$ respectively. However, the SNe in NSWDs and BHWDs occur before mass transfer can significantly tighten systems, such that the median separations at the moment of SN are $\sim 23.7 \rsun$ and $\sim 71.2\rsun$ respectively. As a result, the typical kick applied to DCOs containing a WD is weaker.

\subsection{\lisa detections are mostly insensitive to position}\label{sec:insensitivity-to-position}

The number of \lisa detections is mostly insensitive to the change in position due to SN kicks \citep[in agreement with][]{Storck+2023:2023MNRAS.521.2368S}, despite the large changes in scale height. In Figure~\ref{fig:n_detect} we show the number of \lisa-detectable DCOs for each DCO type and model variation, when using the formation location in the \snr calculation (open points) and when using the present-day position (closed points). We find that the median number of detections decreases by no more than ${\sim}25\%$ for any DCO type, which is significantly smaller than both the uncertainty on each point and the differences between model variations.

The slight decrease in detection rate is due to the kinematic heating of the DCO orbits by SN kicks. Kicked DCOs typically travel further from the Earth on more radial orbits, increasing their distance from \lisa at present day. The \snr of sources scales inversely with distance and therefore accounting for kicks typically reduces the detectability of (more distant) kicked DCOs. The importance of accounting for kicks on the detection rate of each DCO type follows the trends in Figure~\ref{fig:dco_positions}. BHNSs are the most displaced around the Galaxy relative to their formation location and similarly have the largest decrease in detection rate when properly accounting for their Galactic orbits. In contrast, \lisa-detectable NSWDs and BHWDs receive weaker kicks and their detection rates vary by less than $10\%$.

Overall, we find that around 20--240 BH and NS binaries are detectable by \lisa in our models with an \snr above 12 over 8 years of observations. This detection total is slightly lower than some prior works, but we enforce a more stringent detection requirement of \snr$>12$ while earlier works have typically used 7 or 8 \citep[e.g.,][]{COSMIC, Lau+2020, Wagg+2022}. The maximum number of detections is achieved under the optimistic CE assumption \citep[in agreement with][]{Wagg+2022}, because this variation allows binaries to survive CEs initiated by donors without a distinct core-envelope boundary, significantly reducing the number of stellar mergers. Varying the efficiency of common envelope lower or higher both reduce the number of detections \citep{Wagg+2022}, since binaries either merge too often in CEs, or finish a CE on an orbit that is too wide to merge in a Hubble time. Applying the \citet{Maltsev+2025:2025AA...700A..20M} prescription for remnant masses significantly reduces the number of BHs that are formed \citepalias{Wagg+2026:2026arXiv260722814W}, and as such, the number of BHBHs and BHWDs that are detected decreases significantly. We find that the number of detections is not particularly sensitive to variations in the treatment of ECSNe \citep[in contrast to][]{Wagg+2022} or mass transfer stability. We will consider how these variations affect how many BH and NS binaries can be \textit{identified} in Section~\ref{sec:distinguishers}.

\section{Methods for distinguishing NS and BH DCOs from WDWDs in \lisa}\label{sec:dist_methods}

In this Section we consider several potential avenues for distinguishing DCOs containing a NS or a BH from the abundant WDWD \lisa foreground. These DCOs are distinct from WDWDs in many ways; WDWDs are typically less massive, circular, and kinematically similar to the Galactic disc. First, in \S\ref{sec:potential_methods}, we outline several promising source properties that could be used to distinguish DCOs from WDWDs. In \S\ref{sec:intrinsic-props}, we examine these intrinsic properties of \lisa-detectable DCOs. Next, in \S\ref{sec:uncertainties}, we consider how measurement uncertainties impact the utility of each source property. Finally, in \S\ref{sec:distinguishers}, we explore how effective each property is at distinguishing DCOs containing NSs and BHs from WDWDs.

\subsection{Relevant \lisa source properties}\label{sec:potential_methods}


\paragraph{Chirp mass} A common, and intuitive, distinguishing factor between WDWDs and DCOs containing a NS or BH is the chirp mass \citep[e.g.,][]{Nelemans+2001, Belczynski+2008,Lau+2020, Wagg+2022}. The larger mass of BHs and NSs means that DCOs containing these objects typically have a larger chirp mass. In particular, there is an upper limit on the chirp mass of a WDWD, because neither WD can exceed the Chandrasekhar limit ($M_{\rm Ch} \approx 1.44 \msun$). However, measuring the chirp mass precisely requires the detection of a frequency derivative, which is only possible for a subset of more massive DCOs, because they are often stationary on the timescale of the \lisa mission.

\paragraph{Eccentricity} WDWDs are expected to be circular\footnote{Observationally, there are two examples of short-period WDWD binaries with nonzero eccentricity \citep{vanRoestel+2025:2025arXiv250515580V}. However their eccentricities are extremely small $e\approx2\times10^{-3}$ and we calculate (with \legwork) that these systems would circularise to $e \approx 2\times10^{-4}$ before they are resolvable by \lisa.} as a result of mass transfer during their formation \citep[e.g.,][]{Nelemans+2001} or circularisation from GWs. Moreover, recent work has shown that, though many \lisa-detectable WDWDs may form in triples and retain a tertiary companion, no system retains a measurable eccentricity in the \lisa band \citep{Rajamuthukumar+2025:2025A&A...704A.156R}. In contrast, core-collapse events during BH and NS formation often impart significant eccentricity on DCO orbits. As a result, a Galactic \lisa source resolved with measurable eccentricity is likely to host a BH or NS \citep[e.g.,][]{Lau+2020, Andrews+2020:2020ApJ...892L...9A, Wagg+2022, Storck+2023:2023MNRAS.521.2368S, Korol+2024:2024MNRAS.530..844K}.

\paragraph{Distance from the Galactic plane} WDWDs are expected to follow the stellar distribution in the Galaxy for the same reason that they are expected to be circular: WDWDs do not experience core-collapse events. In contrast, DCOs with BHs and NSs are often displaced from the Galactic plane by SN kicks (Figure~\ref{fig:dco_positions}, see also \citealt{underworld}, \citetalias{Wagg+2026:2026arXiv260722814W}). As a result, it has been argued that, if \lisa localises a DCO far enough from the Galactic plane, it is likely to contain a BH or NS \citep[e.g.,][]{Wagg+2022}.

\paragraph{Orbital frequency} The orbital frequency also holds somewhat subtle imprints of the (more poorly measured) chirp mass. The horizon distance\footnote{The maximum distance at which the \snr of a source exceeds the resolvable threshold.} of monochromatic, circular sources scales with chirp mass and orbital frequency as $D_{\rm hor} \propto \mc^{5/3} f_{\rm orb}^{2/3}$ \citep[e.g.,][Eq.~27, 65, and 66]{LEGWORK}. Given that WDWDs are lower mass, this imposes a more stringent selection against low-frequency sources \citep[e.g.,][]{Sesana+2020}. For example, a WDWD with $m_1 = m_2 = 0.6 \msun$ and $f_{\rm orb} = \num{2e-4}\unit{Hz}$ has a horizon distance of $190\unit{pc}$ (under the assumption of an 8-year observation time and a \snr threshold of 12). In contrast, a BHBH with $m_1 = m_2 = 10 \msun$ at the same frequency can be detected out to $20.7\unit{kpc}$, a factor of ${\sim}10^{6}$ increase in volume. Similarly, a typical BHBH in our simulations has median parameters of $\mc = 7.7\msun$, $e = 0.01$, and $D = 11.0 \unit{kpc}$, which produces an \snr of 50.7 at $f_{\rm orb} = \num{3e-4}\unit{Hz}$. A WDWD with $m_1 = m_2 = 0.6\msun$ would need to be at $123\unit{pc}$ to reach the same \snr at this frequency. As a result of this selection, low-frequency sources can only be WDWDs if they are located extremely close to the Earth. Moreover, unlike the chirp mass, the orbital frequency is directly measured by \lisa and often has very small uncertainties (less than $10^{-5}$ relative error for most \lisa-detectable DCOs, see Eq.~\ref{eq:f_orb_unc}) even in the absence of a frequency derivative.

\subsection{Intrinsic \lisa source properties}\label{sec:intrinsic-props}

\begin{figure*}
    \centering
    \includegraphics[width=\textwidth]{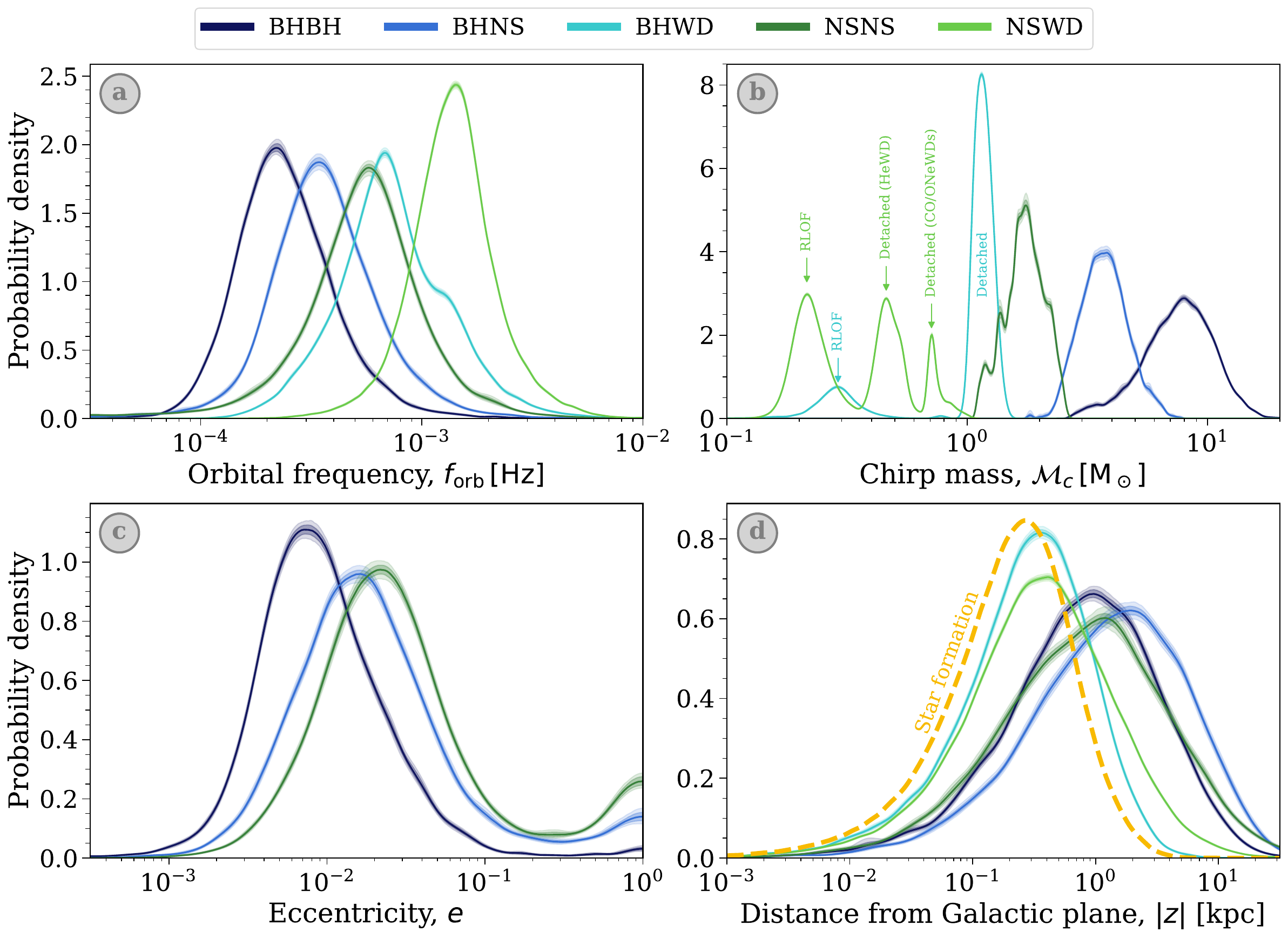}
    \caption{DCOs containing BHs and NSs are distinct from WDWDs across several properties. Each panel shows the probability density of a different property for the fiducial population of \lisa-detectable DCOs in a 8-year mission for the fiducial model. The distributions are shown with kernel density estimators (KDEs), with the 1- and 2-$\sigma$ uncertainty indicated by the shaded regions. Note that these distributions do not account for measurement uncertainties, see Figure~\ref{fig:dco-measured-props}. \textbf{Panel a} shows the orbital frequency of each DCO type at the time of the \lisa mission. \textbf{Panel b} shows the chirp mass, with several annotations that explain the multimodal distributions. \textbf{Panel c} shows the eccentricity, excluding BHWDs and NSWDs since we find that 100\% of those DCOs are circular by present day. \textbf{Panel d} shows the distance from the Galactic plane, with a dashed yellow line to show the distribution of star formation, which should approximate the distribution of WDWDs.}
    \label{fig:dco-props}
\end{figure*}

In Figure~\ref{fig:dco-props} we show the intrinsic distribution of the relevant properties for the \lisa-detectable population of DCOs (assuming an 8-year observation duration and detectable threshold of \snr $> 12$).

The orbital frequencies of BH and NS DCOs are typically lower than WDWDs (Figure~\ref{fig:dco-props}a), with more massive DCOs being skewed to lower frequencies. As discussed in Section~\ref{sec:potential_methods}, DCOs with higher masses can be detected to larger distances at a given frequency. DCOs spend a much larger fraction of their lifetimes at lower frequencies because their inspiral rate scales strongly with frequency \citep[$\dot{f} \propto f_{\rm orb}^{11/3}$, see e.g.,][Equation~25]{LEGWORK}. Therefore, more massive DCOs, such as BHBHs, are much more often detected at lower frequencies.

The chirp mass distributions of different DCO types are distinct and often separable (Figure~\ref{fig:dco-props}b). NSWDs and BHWDs have multimodal distributions due to the contributions of different subpopulations. Older systems have inspiralled enough to have reached RLOF, during which the WD donates mass to the compact object companion, decreasing in mass and producing the lowest chirp mass subpopulations. The detached systems for NSWDs can be further separated between those systems containing helium WDs as opposed to more massive carbon-oxygen and oxygen-neon WDs. The vast majority of BHBHs and BHNSs (and a large subset of NSNSs) have chirp masses above the maximum possible WDWD mass, although confidently measuring these masses can be difficult (see Section~\ref{sec:uncertainties}).

DCOs not containing a WD always exhibit at least mild eccentricity ($e \sim 0.01$), with some rarer high eccentricity ($e > 0.5$) subpopulations (Figure~\ref{fig:dco-props}c). SN kicks drive this eccentricity and these are often stronger for systems containing a NS (see Figure~\ref{fig:kick_mags}), thus those systems are more eccentric. Gravitational wave emission circularises the binaries over time \citep{Peters+1964} and as such we find that \lisa DCOs with $e > 0.5$ were formed less than 100 Myr ago. We find that \lisa detectable NSWDs and BHWDs are circular by present day. Although these systems experienced a core-collapse event that may have driven eccentricity, subsequent mass transfer episodes have circularised the systems. These episodes either occur during the evolution of the secondary star, or after DCO formation as a result of gravitational-wave inspiral.

DCOs that contain a BH or NS are distributed further from the Galactic plane than the stellar distribution (Figure~\ref{fig:dco-props}d). As we highlighted in Figure~\ref{fig:dco_positions}, the present-day positions of these DCOs are dispersed far from the Galactic plane by SN kicks. The differences from the stellar distribution (shown in gold) are particularly stark for DCOs not containing WDs, which can survive stronger kicks from the secondary star (Figure~\ref{fig:kick_mags}).

\subsection{Accounting for measurement uncertainties}\label{sec:uncertainties}

\begin{figure*}
    \centering
    \includegraphics[width=\textwidth]{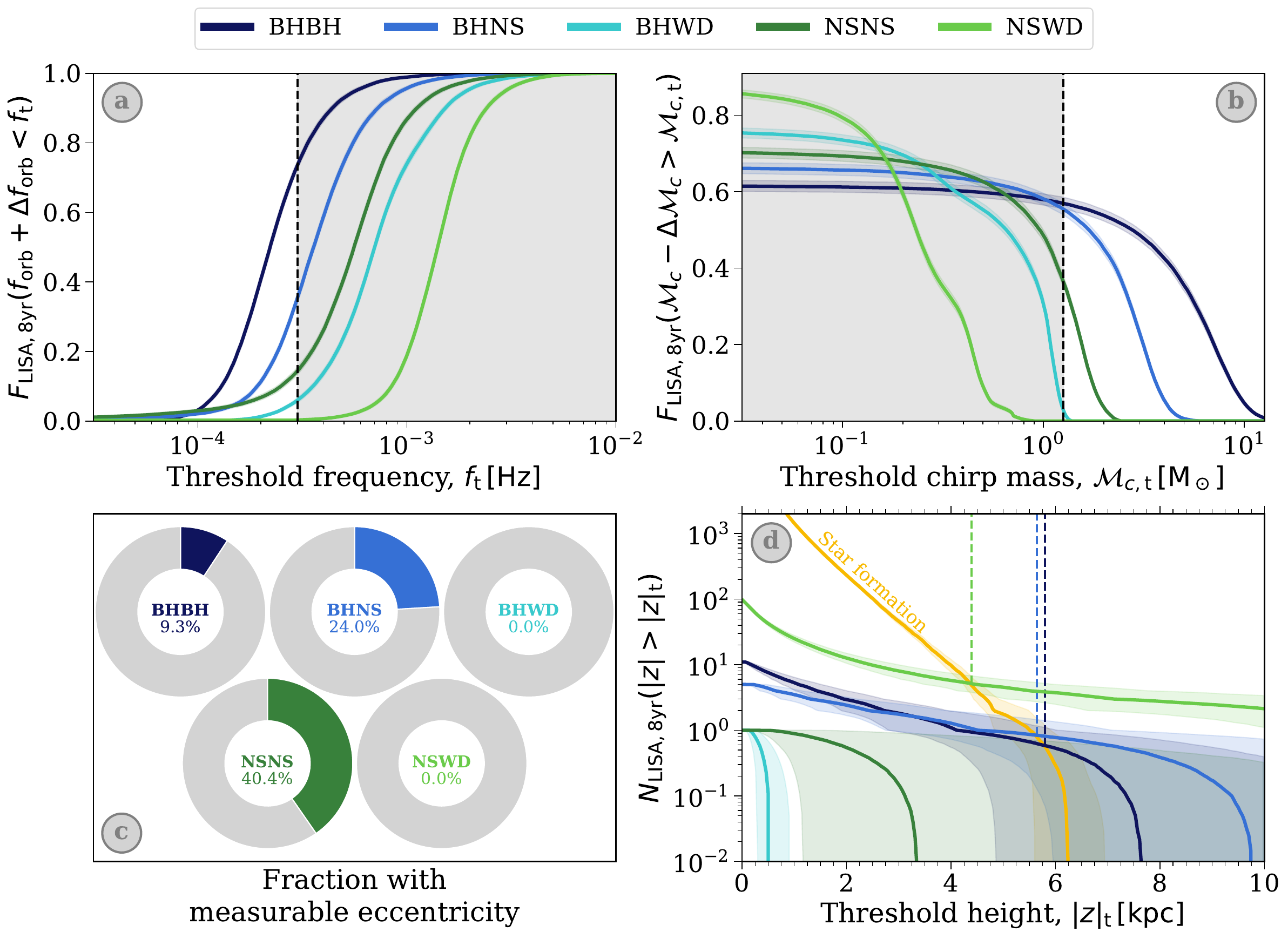}
    \caption{When accounting for measurement uncertainties, distinguishing from the WDWD population is challenging. \textbf{Panel a} shows the cumulative fraction of \lisa-detectable sources with $f_{\rm orb} + \Delta f_{\rm orb}$ below a given frequency for each DCO type. The dashed line indicates the minimum frequency at which a WDWD is expected to be detected \citep{Korol+2022:2022MNRAS.511.5936K, Thiele+2023:2023ApJ...945..162T}. \textbf{Panel b} shows the cumulative fraction of \lisa-detectable sources with $\mc - \Delta \mc$ above a given mass for each DCO type. Many sources have no measureble chirp mass and thus the curves do not start at 1. The dashed line indicates the maximum possible mass of a WDWD. \textbf{Panel c} contains pie charts showing the fraction of DCOs that have measurable eccentricity either because (a) two individual harmonics are resolvable, or (b) the sources eccentricity exceeds $e_{\rm min}(f_{\rm dom}, \rho)$ following \citet{Moore+2024:2024MNRAS.531.2817M}. \textbf{Panel d} shows the number of \lisa-detectable DCOs above a given distance from the Galactic plane. The gold line approximates the WDWD distribution, with 25000 detections and following the stellar distribution. Dashed lines indicate the height at which each DCO type dominates over WDWDs. All panels use the fiducial model.}
    \label{fig:dco-measured-props}
\end{figure*}

Importantly, \lisa cannot measure all of these distinguishing properties perfectly and as such, the uncertainty on these values determines how many sources can be confidently distinguished from the WDWD population. We estimate the uncertainties on each source property in Appendix~\ref{app:unc_estimates}. The most precisely measured quantity is the orbital frequency of the source. Given that WDWDs cannot be resolved at low frequency due to their lower masses, one can define a criterion based on orbital frequency.
\begin{equation}\label{eq:f_crit}
    f_{\rm orb} + \Delta f_{\rm orb} < f_{\rm orb}^{\rm WDWD, min}, \quad \text{($f_{\rm orb}$ criterion)}
\end{equation}
where $\Delta f_{\rm orb}$ is given in Eq.~\ref{eq:f_orb_unc} and we assume $f_{\rm orb}^{\rm WDWD, min} = \num{3e-4}\unit{Hz}$ based on population synthesis of WDWDs \citep{Korol+2022:2022MNRAS.511.5936K, Thiele+2023:2023ApJ...945..162T}. A circular WDWD with $m_1 = m_2 = 0.6 \msun$ at this frequency is resolvable up to around $350\unit{pc}$ from Earth (compared to $39\unit{kpc}$ for a BHBH with $m_1 = m_2 = 10\msun$). In Figure~\ref{fig:dco-measured-props}a, we show the cumulative fraction of \lisa-detectable DCOs with orbital frequencies measured below a given frequency. The black dashed line indicates $f_{\rm orb}^{\rm WDWD, min}$ and therefore the intercept with this dashed line indicates the fraction that can be separated from WDWDs. In the fiducial model, this is extremely effective for BHBHs (${\sim}70\%$) and still useful for BHNS and NSNS.

A DCO containing a BH or NS can be identified if its chirp mass is confidently above the maximum possible chirp mass of a WDWD. A WDWD for which both components are at the Chandrasekhar limit, such that $m_1 = m_2 = 1.44 \msun$, has a chirp mass of $\mc^{\rm WDWD, max} \approx 1.25 \msun$. Therefore, a DCO can be distinguished from the WDWD foreground if the following criterion is met
\begin{equation}\label{eq:mc_crit}
    \mc - \Delta \mc > \mc^{\rm WDWD, max} \quad \text{($\mc$ criterion)}
\end{equation}
We calculate the chirp mass uncertainty of each source using Eq.~\ref{eq:chirp_mass_uncertainty} and show the cumulative fraction of DCOs with masses measured above a given threshold mass in Figure~\ref{fig:dco-measured-props}b. $\mc^{\rm WDWD, max}$ is indicated by a dashed line and therefore the intercept of each curve with this line gives the fraction of systems distinguishable by mass alone. Around $60\%$ of BHBHs and BHNSs, and $40\%$ of NSNSs can be distinguished with this method. BHWDs and NSWDs are almost always too low mass to be distinguishable based on chirp mass. We highlight that the chirp mass cannot be measured at all for a significant fraction of each DCO type (as shown by the $y$-intercept of each curve).

The measurement of eccentricity in \lisa relies on the identification and comparison of different gravitational-wave harmonics. Circular sources emit at only the $n = 2$ harmonic, but eccentric sources produce emission at potentially many harmonics. We follow the same method as \citet{Korol+2024:2024MNRAS.530..844K} to determine which sources have a measurable eccentricity based on two criteria: either (a) the source has two harmonics that are each individually resolvable by \lisa or (b) the eccentricity of the source exceeds the minimum detectable eccentricity calculated by \citet{Moore+2024:2024MNRAS.531.2817M}. To summarise, we consider a source to have a measurable eccentricity (and thus not be a WDWD) if
\begin{equation}\label{eq:e_crit}
    e > e_{\rm min}(f_{\rm dom}, \rho) \text{ or } N_{\rm harm} \ge 2, \quad \text{($e$ criterion)}
\end{equation}
where $e_{\rm min}(f_{\rm dom}, \rho)$ is given by Equation 21 of \citet{Moore+2024:2024MNRAS.531.2817M} and $N_{\rm harm}$ is the number of individually detectable harmonics. In Figure~\ref{fig:dco-measured-props}c, we show the fraction of each DCO type that has a measurable eccentricity. The fit from \citet{Moore+2024:2024MNRAS.531.2817M} is applied only in the tested frequency regime ($f_{\rm dom} > 0.5 \unit{mHz})$. Unsurprisingly, the circular NSWDs and BHWDs (see Figure~\ref{fig:dco-props}) have no measurable eccentricity. As one may expect from Figure~\ref{fig:dco-props}c, the measurable fraction increases from BHBHs to BHNSs to NSNSs, reaching $40\%$ for NSNSs.

Finally, we consider how finding sources far from the Galactic plane may distinguish them from WDWDs. From Figure~\ref{fig:dco-props}d, one may conclude that finding a source at a distance of more than $1\unit{kpc}$ would make it far more likely to be a DCO containing a BH or NS. However, the normalisation of these curves is critical. Given that the Milky Way stellar discs follow exponential distributions, finding sources far from the Galactic plane can still occur rarely. Therefore, even though a small \textit{fraction} of WDWDs will be detected far from the plane, the \textit{number} of detections still remains larger than other DCOs until one looks much further from the plane. We highlight this in Figure~\ref{fig:dco-measured-props}d, which shows the cumulative number of DCOs detectable above a given distance from the Galactic plane. For this Figure we assume that WDWDs are distributed in the same manner as stars in the Galaxy and that 25000 are detected by \lisa \citep[e.g.,][]{Korol+2017}. The dashed lines indicate the heights at which each DCO type dominates over the WDWD population. NSWDs eclipse the WDWDs at around $|z|_t = 4.4 \unit{kpc}$, while the BHBH and BHNSs require $|z|_t \sim 5.7 \unit{kpc}$. In contrast, NSNS and BHWDs cannot be distinguished from WDWDs via position in this manner. In reality, the WDWD population is likely skewed towards slightly younger (and kinematically cooler) ages than the general stellar population \citep[e.g.,][]{Thiele+2023:2023ApJ...945..162T} and thus may be distributed closer to the plane. We assume that a DCO detected by \lisa fulfilling the following criterion is distinguishable from WDWDs
\begin{equation}\label{eq:z_crit}
    |z| - \Delta |z| > |z|_t,\quad\text{($|z|$ criterion)}
\end{equation}
where $\Delta |z|$ is the 1-$\sigma$ uncertainty on $|z|$ (Eq.~\ref{eq:z_unc}) and $|z|_t$ is the threshold height determined by the intercept of the CDFs shown in Figure~\ref{fig:dco-measured-props}d. We find that only a small fraction of DCOs meet this criterion: 3\% of BHBHs, 9\% of BHNSs, and 2\% of NSWDs. Although these fractions are not useful for the rare BHBHs and BHNSs, given the higher detection rate of NSWDs and weakness of other techniques for these DCOs, this may be the most effective strategy for identifying a handful of NSs within NSWD systems.


\subsection{Efficacy of different strategies for distinguishing from WDWDs}\label{sec:distinguishers}

\begin{figure*}
    \centering
    \includegraphics[width=\textwidth]{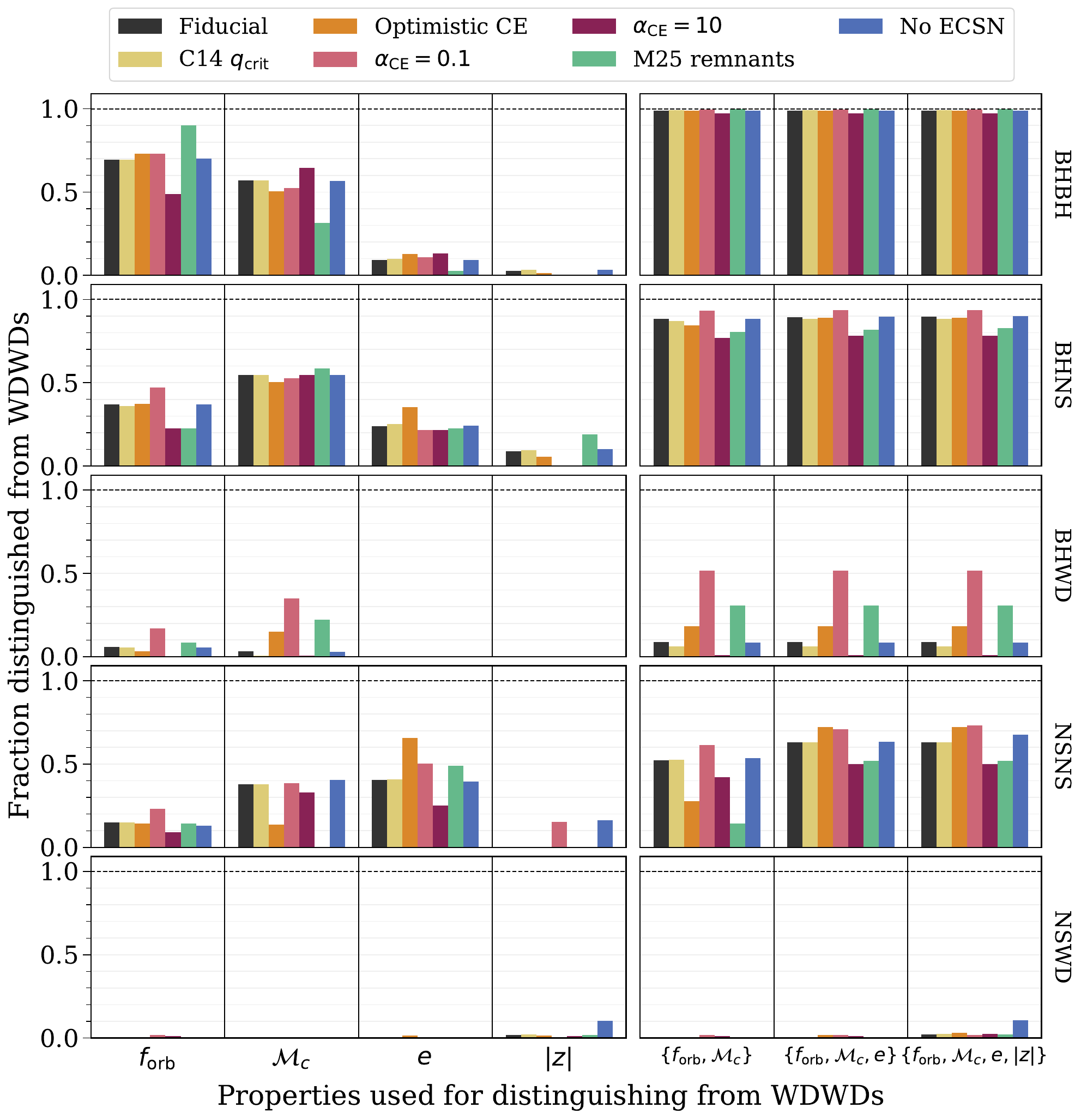}
    \caption{Some \lisa source properties are more effective at distinguishing DCOs with BHs and NSs from WDWDs. In particular, most BHBHs and BHNSs can be distinguished from WDWDs by frequency and chirp mass alone, while NSNS identification is aided by eccentricity measurements. Each column shows the fraction of DCOs that can be distinguished from WDWDs when using different source properties. $f_{\rm orb}$ refers to systems with an orbital frequency measured below $f_{\rm orb}^{\rm min, WDWD} = 3\times10^{-4} \, {\rm Hz}$ (Eq.~\ref{eq:f_crit}), $\mc$ to systems with a chirp mass measured above $\mc^{\rm WDWD, max} = 1.25 \msun$ (Eq.~\ref{eq:mc_crit}), $e$ is systems with measurable eccentricity (Eq.~\ref{eq:e_crit}), and $|z|$ is systems beyond the height from the Galactic plane at which we expect those DCOs to outnumber WDWDs (Eq.~\ref{eq:z_crit}). The last three columns combine unions of these criteria. Each row corresponds to a different DCO type (annotated on the right) and each bar represents a different model variation.}
    \label{fig:distinguishers}
\end{figure*}

In this Section we consider the efficacy of the criteria defined in the prior Section, both separately and in combination, for distinguishing \lisa DCOs containing BHs and NSs from WDWDs. In Figure~\ref{fig:distinguishers} we show the fraction of DCOs that can be distinguished from WDWDs via different strategies. The first four columns show the fraction distinguished when applying an individual criterion, while the last three combine multiple criteria. Each row shows a different DCO type and the coloured bars indicate different model variations (see Section~\ref{sec:cosmic_explained}), and as such the black bars in the first four columns give the values discussed in the prior Section.

The orbital frequency and chirp mass criteria distinguish complementary subpopulations. The $\{f_{\rm orb}, \mc\}$ column in Figure~\ref{fig:distinguishers} shows that the combination of these criteria is particularly effective, in that there are few sources that were distinguished by both $f_{\rm orb}$ \textit{and} $\mc$. This complementarity occurs because precise chirp mass measurements, necessary to assert the DCO is too massive to be a WDWD, require a measurable frequency derivative, which is much stronger at higher frequencies ($\dot{f} \propto f^{11/3}$). In contrast, distinguishing based on orbital frequency is only effective at low frequencies. As a result, distinguishing based on orbital frequency is an effective method for finding DCOs that would be missed when only applying chirp mass or eccentricity criteria.

The addition of eccentricity as a further criterion is primarily helpful for identifying NSNS binaries. The $\{f_{\rm orb}, \mc, e\}$ column in Figure~\ref{fig:distinguishers} shows that the distinguished fraction is almost unchanged for each DCO type except NSNSs. Although BHBH and BHNS binaries have sizeable subpopulations that have measurable eccentricities (see Figure~\ref{fig:dco-measured-props}c), these subpopulations are already distinguished based on their orbital frequency or chirp mass. For NSNSs, which typically have lower masses and higher orbital frequencies, their significantly eccentric subset are effective at distinguishing them from WDWDs.

Separating from WDWDs based on distance from the Galactic plane provides negligible distinguishing power in addition to other methods except in the case of NSWDs. From a comparison of the $\{f_{\rm orb}, \mc, e\}$ and $\{f_{\rm orb}, \mc, e, |z|\}$ columns in Figure~\ref{fig:distinguishers}, one can note that sources have almost always been identified by another criterion without accounting for $|z|$. The exception to this is NSWD binaries, which are not distinguishable by other means. Despite the small fraction ($\sim 2$--3\%), this identifies around 2 NSWDs in the fiducial model, and in the variation without ECSNe this increases to around 11 NSWDs. Therefore, this can be a fruitful source of NSs that is comparable to the more identifiable but rarer BHNSs and NSNSs.

The distinguishable fraction is mostly robust to variations in binary stellar evolution. For BHBHs and BHNSs, the fractions vary between models by at most around 10\%. NSNSs and BHWDs are slightly more sensitive to variations, but in both cases the number of detections is low so this sensitivity is not particularly important. For NSWDs, the fractions are quite consistent except in the case without ECSNe. In this model, progenitors from the fiducial model that would explode in ECSNe instead undergo core-collapse SNe, receiving larger kicks and moving further from the Galactic plane (and hence are more distinguishable based on $|z|$). Therefore, overall the uncertainty in number of BH and NS binaries that can be identified from assumptions in binary evolution is driven primarily by the total number of detections, rather than the fraction that can be distinguished.

\section{Discussion} \label{sec:discussion}

\subsection{What can we learn from BH and NS binaries identified via orbital frequency alone?}\label{sec:dcos_without_masses}

Combining the orbital frequency and chirp mass criteria increases the fraction of BHBHs and BHNSs that can be distinguished from WDWDs from 50--60\% to 90--100\% (Section~\ref{sec:distinguishers}). However, the additional sources identified by orbital frequency are predominantly those without a measurable frequency derivative, and hence without a measured chirp mass. Identifying a source in this way therefore only tells us that it likely hosts a BH or NS. It reveals neither its component masses nor even which type of DCO it is.

Nevertheless, the number of identified BH and NS binaries is itself a sensitive probe of binary evolution. The number of \lisa-detectable DCOs varies substantially between model variations \citep[Figure~\ref{fig:n_detect}, see also e.g.,][Figure 8]{Wagg+2022}, whereas the fraction of these that can be distinguished from WDWDs is comparatively robust (Section~\ref{sec:distinguishers}). The number of identified BH and NS binaries therefore closely tracks the underlying detection rate.

At the same time, sources identified by orbital frequency still carry some information about their masses. For a stationary source, the measured strain amplitude constrains a combination of chirp mass and distance ($\mc^{5/3}/D$). A prior on the Galactic distribution of sources, which disfavours very small distances, therefore yields a probabilistic lower bound on the chirp mass; this is the same reasoning that underlies our orbital frequency criterion. Conversely, the absence of a measurable frequency derivative (Eq.~\ref{eq:f_orb_dot_unc}), together with the finite extent of the Galaxy, implies a potential upper bound, though these DCOs could be detected beyond the Galaxy \citep[e.g.,][]{Korol+2020, Keim+2023:2023MNRAS.521.1088K}. These bounds could be incorporated into a hierarchical inference of the DCO mass distribution, even when no individual source has a well-measured mass. Moreover, \citet{Seto+2001:2001PhRvL..87y1101S} proposed that, for eccentric binaries detected at low frequencies in \lisa without a frequency derivative, one may determine the total mass of the DCO via a signature of periastron advance \citep[see however,][for limitations in applying this to systems with BHs]{Seto+2016:2016MNRAS.460L...1S}. Finally, since DCOs containing NSs and BHs are not expected to have significant EM signatures (except in the case of young pulsars), EM searches consistent with the horizon distance of WDWDs or NSWDs at a given frequency could prove to be especially fruitful in discovering a binary \citep[e.g.][]{Tauris+2018:2018PhRvL.121m1105T,Korol+2026:2026A&A...705A.154K} or ruling out a low-mass system. 

\subsection{Negative frequency derivatives could identify NSWDs undergoing RLOF}\label{sec:negative_chirps}

NSWDs and BHWDs are difficult to distinguish from WDWDs with the methods we have explored thus far (Figure~\ref{fig:distinguishers}), yet those undergoing RLOF may be identifiable by their negative frequency derivatives. These DCOs will first inspiral due to GWs until the WD overflows its Roche lobe, initiating mass transfer onto its companion. Due to the mass ratio of the system, this mass transfer will cause the binary to widen and produce a negative frequency derivative \citep[e.g.,][]{Kremer+2017:2017ApJ...846...95K, Tauris+2018:2018PhRvL.121m1105T}, which continues as the WD expands due to mass loss.

We find that around 60\% of NSWDs, and 20\% of BHWDs, are undergoing RLOF when detected by \lisa in the fiducial model. Accreting WDWDs, which also have negative frequency derivatives, are relatively rarer within the \lisa-detectable WDWD population, representing around 10\% of detections \citep{Kremer+2017:2017ApJ...846...95K}. Therefore, the ratio of accreting NSWDs to accreting WDWDs is much more favourable than total number of NSWDs to WDWDs in \lisa. Given the rarity of \lisa-detectable BHWDs, the 20\% undergoing RLOF are still unlikely to be distinguished from WDWDs, thus we consider only NSWDs below.

Most accreting NSWDs are detected shortly after the onset of RLOF. This selection makes sense, since these DCOs produce the strongest gravitational-wave emission shortly after the onset of RLOF. Gravitational-wave emission scales with the chirp mass and frequency of the system (assuming its distance and eccentricity do not change significantly). The chirp mass can be written as $\mc = M_{\rm tot} \, \eta^{3/5}$, where $M_{\rm tot}$ is the total mass and $\eta$ is the symmetric mass ratio. Therefore, as the mass transfer proceeds, both the frequency and chirp mass decrease as the system widens and as the mass ratio becomes more extreme, respectively. As a result, the peak gravitational-wave emission occurs shortly after the onset of RLOF, as the binary begins to expand.

A large fraction of accreting NSWDs have measurably negative frequency derivatives. Accreting NSWDs will widen more slowly over time, and thus the strongest negative frequency derivatives occur shortly after mass transfer begins, which is also when they most strongly emit gravitational waves. In the fiducial model, we find that ${\sim}50\%$ of accreting NSWDs (${\sim}30\%$ of all \lisa-detectable NSWDs) have measurable negative frequency derivatives ($\Delta \dot{f} / \dot{f} < 1$). This corresponds to around 30 NSWDs detected by \lisa for this model.

Combining a measurement of a negative frequency derivative with a sky localisation in \lisa could prove fruitful for identifying Galactic NSWDs. Firstly, given that accreting WDWDs are much rarer than detached WDWDs \citep{Kremer+2017:2017ApJ...846...95K}, one could apply a more favourable $|z|_{t}$ (Eq.~\ref{eq:z_crit}) for identifying accreting NSWDs. Moreover, with a precise sky localisation, one could search for EM counterparts to these \lisa detections, which could be observable as ultra-compact X-ray binaries \citep[e.g.,][]{Tauris+2018:2018PhRvL.121m1105T}, which may differentiate between the DCO types.
Overall, although NSWDs are difficult to identify by their frequency, chirp mass, eccentricity, or position, their frequency \textit{derivatives} seem promising for expanding our observations of Galactic neutron stars.

\subsection{Caveats}\label{sec:caveats}

Our predictions are subject to several uncertainties, both in terms of binary population synthesis and \lisa data analysis.
We highlight some main concerns that are particularly relevant to our results.

\paragraph{Minimum WDWD frequency} Our orbital frequency criterion (Eq.~\ref{eq:f_crit}) relies on a threshold frequency below which WDWDs are not expected to be resolvable, which we take from population synthesis of the Galactic WDWD population \citep{Korol+2022:2022MNRAS.511.5936K, Thiele+2023:2023ApJ...945..162T}. This is not a strict physical limit. A sufficiently nearby WDWD could, in principle, be resolved at lower frequencies, and the precise threshold depends on the assumed WDWD population model and local space density. Our results are thus sensitive to the choice of frequency; setting $f_{\rm orb}^{\rm WDWD, min}$ to a lower threshold of $2\times10^{-4}\unit{Hz}$ reduces the fraction of systems identified by frequency alone to 0.37 and 0.12 for BHBHs and BHNSs respectively in the fiducial model (compared to 0.69 and 0.37 with $f_{\rm orb}^{\rm WDWD, min} = \num{3e-4}\unit{Hz}$). However, it is worth noting that the distance to which a WDWD can be resolved declines steeply towards lower frequencies, owing to both the decreasing strain amplitude and the degrading sensitivity of \lisa. At our adopted threshold a typical WDWD is resolvable only within ${\sim}350\unit{pc}$. This volume shrinks rapidly at lower frequencies, so the expected number of contaminating WDWDs drops quickly below the threshold.

\paragraph{\snr calculations for NSWDs and BHWDs undergoing RLOF} We compute the \snr of all NSWDs and BHWDs under the assumption that they are stationary sources. Although we find that the stationary approximation is a valid choice for all accreting NSWDs and BHWDs in our sample, as discussed in Section~\ref{sec:negative_chirps}, many of these systems are evolving slowly towards lower frequencies due to RLOF. Dedicated searches with templates that allow for $\dot{f} < 0$ may be required to recover these sources \citep[e.g.,][]{Kremer+2017:2017ApJ...846...95K, Tauris+2018:2018PhRvL.121m1105T}. As such our predictions for the detection rates of NSWDs and BHWDs may be overestimated if these searches are not performed.

\paragraph{Stability of mass transfer in NSWDs} The stability of mass transfer from a WD onto a NS remains uncertain. In our simulations, we determine mass transfer stability based on critical mass ratios. For WDs transferring mass onto a NS, we assume mass transfer is stable as long as the ratio of the WD to the NS is $q = m_{\rm WD} / m_{\rm NS} < 0.628$. However, hydrodynamical work suggests that only NSWDs with WD masses below ${\sim}0.2\msun$ undergo stable mass transfer, with more massive WDs instead undergoing unstable mass transfer and being disrupted \citep{Bobrick+2017:2017MNRAS.467.3556B}. This limit is significantly lower than our assumption; for a typical NS mass of $1.4 \msun$, we would set a maximum WD mass based on the mass ratio at $0.88\msun$. We find that only around 8\% of the accreting NSWDs in our fiducial model initiate mass transfer with $m_{\rm WD} < 0.2 \msun$. Assuming the other 92\% of systems instead merge would reduce our predicted NSWD detections by a factor of 2. Many of these WDs only slightly exceed this limit, such that $70\%$ have $m_{\rm WD} < 0.3 \msun$. The number of NSWD detections in \lisa is therefore highly sensitive to this limit and, though it increases the uncertainty on our results, this implies \lisa may offer constraints on the stability of mass transfer from WDs onto NSs.

\paragraph{Associating individually detected harmonics} Our eccentricity criterion (Eq.~\ref{eq:e_crit}) considers a source measurably eccentric if at least two of its harmonics are individually resolvable. However, a search using quasi-circular templates would recover these harmonics as separate, circular sources, which must then be associated with one another. This association should be possible, as \citet{Moore+2024:2024MNRAS.531.2817M} demonstrate with a detection threshold of $\rho \ge 6$. It is likely that \lisa global fit pipelines that consider multiple source types \citep[e.g.,][]{Littenberg+2023:2023PhRvD.107f3004L} will be able to correctly associate harmonics, especially given our more conservative threshold for individual harmonics of $\rho \ge 12$.

\paragraph{Approximate uncertainty estimates} Our measurement uncertainties for \lisa sources follow approximate analytic scalings (Appendix~\ref{app:unc_estimates}). In reality, the global fit must instead simultaneously characterise tens of thousands of overlapping Galactic sources. As a result, parameter correlations and confusion with overlapping sources may degrade the uncertainties for some sources. We expect that this would most strongly affect the chirp mass and $|z|$ criteria, which depend on parameters that are measured less precisely.

\section{Summary and Conclusions} \label{sec:conclusions}

In this paper we have presented predictions for the \lisa-detectable population of BHBH, BHNS, NSNS, BHWD, and NSWD binaries, using self-consistent population synthesis and galactic dynamics simulations from \cogsworth, which uses \cosmic and \gala. We investigated the impact of Galactic orbit integration on the \lisa-detectable population of DCOs. We considered several strategies for distinguishing DCOs containing BHs and NSs from WDWDs. We show that a combination of orbital frequency, chirp mass, eccentricity, and distance from the Galactic plane can recover a large fraction of these DCOs that would otherwise be missed with a chirp mass cut alone. We highlight some of our main conclusions below.

\begin{conclusions}
    \conclusion{%
        The detectability of DCOs by \lisa is mostly insensitive to a full treatment of Galactic orbit integration
    }{
        The number of \lisa-detectable DCOs does not vary strongly (decreasing slightly) when accounting for present-day positions after integrating Galactic orbits instead of using DCO formation positions (Figure~\ref{fig:n_detect}). However, the detected systems \textit{are} found further from the plane, with scale heights increasing from their formation locations by 2--7$\times$ depending on the DCO type (Figure~\ref{fig:dco_positions}).
    }

    \conclusion{%
        Distinguishing via chirp mass measurements alone cannot identify most DCOs containing BHs and NSs from WDWDs
    }{
        Requiring that the measured chirp mass be confidently above the maximum possible WDWD mass recovers 50--60\% of BHBHs and BHNSs, and ${\sim}40\%$ of NSNSs, but not BHWDs and NSWDs (Section~\ref{sec:uncertainties}).
    }

    \conclusion{%
        \lisa sources detected at low frequencies are unlikely to be WDWDs, aiding BH and NS identification
    }{
        WDWDs cannot be detected at low frequencies ($f_{\rm orb} \lesssim \num{3e-4} \unit{Hz}$) as their lower masses would require them to be located extremely close to the Earth (Section~\ref{sec:potential_methods}). Selecting sources with low orbital frequencies (Eq.~\ref{eq:f_crit}) can recover a large fraction of more massive DCOs, particularly BHBHs (${\sim}70\%$, see Figure~\ref{fig:dco-measured-props}a)
    }

    \conclusion{%
        Orbital frequency identifies a complementary subset of DCOs containing BHs and NSs to those distinguished via chirp mass
    }{
        Chirp mass can only be measured for sources with a significant frequency derivative, which occurs at high frequency. Therefore, the combination of chirp mass and orbital frequency is particularly effective at identifying BH and NS DCOs, recovering 90--100\% of BHBHs and BHNSs, and ${\sim}50\%$ of NSNSs (Section~\ref{sec:distinguishers}).
    }

    \conclusion{%
        Eccentricity is most helpful in identifying NSNS binaries
    }{
        NSNSs are the most eccentric DCO type of those we considered (Figure~\ref{fig:dco-props}c) and hence have the largest fraction of measurably eccentric sources (Figure~\ref{fig:dco-measured-props}c). Additionally distinguishing by eccentricity can help to identify a further 10--20\% of NSNSs. However, other DCO types are either too circular, or are already identified by mass or frequency for eccentricity to be useful.
    }

    \conclusion{%
        Distance from the Galactic plane only aids in the identification of NSWDs
    }{
        Although most BH and NS DCOs are distributed much farther from the plane than most WDWDs (Figure~\ref{fig:dco-props}d), their lower detection rates mean that extremely large distances from the plane are required to identify DCOs in most cases (Figure~\ref{fig:dco-measured-props}d). NSWDs are numerous enough that a small fraction of them can be identified in this manner, yielding a handful of NS detections (Section~\ref{sec:uncertainties}).
    }
\end{conclusions}

\section{Data and code availability}\label{sec:availability}

The simulated sources are available on Zenodo under an open-source Creative Commons Attribution license: \dataset[doi:10.5281/zenodo.23520103]{https://doi.org/10.5281/zenodo.23520103}. The code for reproducing figures and results is available \href{https://github.com/TomWagg/lisa-dco-identification}{on GitHub}\footnote{\url{https://github.com/TomWagg/lisa-dco-identification}}. This code is also archived on Zenodo: \dataset[doi:10.5281/zenodo.23522490]{https://doi.org/10.5281/zenodo.23522490}.

\begin{acknowledgments}
We thank Will Farr for an insightful conversation that led to the intuition for using orbital frequency to distinguish from WDWDs. We thank Stephen Justham for help in ensuring the correctness of our implementation of the \stroop algorithm. We also thank Konstantin Leyde, Adrian Price-Whelan, Jakob Stegmann, and Brendan O'Connor for helpful discussions related to this work. This research was supported in part by grant NSF PHY-2309135 to the Kavli Institute for Theoretical Physics (KITP). FEP acknowledges support from the Summer Undergraduate Research Fellowship (SURF) from the Office of Undergraduate Research and Scholar Development (OURSD) and the Porges Family Fund for Undergraduate Research at Carnegie Mellon University. KB acknowledges support from the Falco DeBenedetti Career Development Professorship, NASA ATP Grant 80NSSC24K0768.
MZ gratefully acknowledges funding from the Brinson Foundation in support of astrophysics research at the Adler Planetarium. 

\end{acknowledgments}

\software{This work made use of the following software packages: \texttt{astropy} \citep{astropy:2013,astropy:2018,astropy:2022,astropy_21262391}, \texttt{Jupyter} \citep{2007CSE.....9c..21P,kluyver2016jupyter}, \texttt{matplotlib} \citep{Hunter:2007}, \texttt{numpy} \citep{numpy}, \texttt{pandas} \citep{mckinney-proc-scipy-2010,pandas_21500199}, \texttt{python} \citep{python}, \texttt{scipy} \citep{2020SciPy-NMeth,scipy_15366870}, \texttt{COSMIC} \citep{COSMIC}, \texttt{Cython} \citep{cython:2011}, \texttt{gala} \citep{Gala,Gala_17890972}, \texttt{h5py} \citep{collette_python_hdf5_2014}, \texttt{legwork} \citep{LEGWORK_joss,LEGWORK,legwork_v1}, \texttt{Numba} \citep{numba:2015,Numba_21895810}, \texttt{schwimmbad} \citep{schwimmbad}, \texttt{seaborn} \citep{Waskom2021}, and \texttt{tqdm} \citep{tqdm_21624223}.
This research has made use of the Astrophysics Data System, funded by NASA under Cooperative Agreement 80NSSC21M00561.
This research made use of \cogsworth and its dependencies \citep{Wagg+2025:2025JOSS...10.7400W,cogsworth_apjs,COSMIC,Gala}. Calculation of \lisa gravitational wave signatures was performed using \legwork \citep{LEGWORK_joss,LEGWORK}.
Software citation information aggregated using \texttt{\href{https://www.tomwagg.com/software-citation-station/}{The Software Citation Station}} \citep{software-citation-station-paper,software-citation-station-zenodo}.}

\disclosure{An LLM (Claude Opus 5.5) was used to proofread the manuscript and correct grammar.}

\begin{contribution}
    T.\ Wagg led the analysis and writing of the paper. F.~E.\ Prince conducted the initial analysis and exploration that was built upon for this work. K.\ Breivik contributed to the writing of the paper and interpretation. Authors 4--6 (K.\ Baasch, J.\ McDermott, and M.\ Zevin) contributed to the interpretation of results.
\end{contribution}

\bibliographystyle{aasjournal}
\bibliography{bibs/paper, bibs/software}{}

\restartappendixnumbering

\allowdisplaybreaks

\appendix{}

\section{Uncertainty estimates in \lisa}\label{app:unc_estimates}

In this Appendix, we provide expressions for estimating the uncertainty on several quantities of \lisa sources. The uncertainty on the orbital frequency can be estimated following \citet{Takahashi+2002}, such that
\begin{equation}\label{eq:f_orb_unc}
    \frac{\Delta f_{\rm orb}}{f_{\rm orb}} = \num{2.5e-6} \qty(\frac{7}{\rho}) \qty(\frac{4\unit{yr}}{T_{\rm obs}}) \qty(\frac{1\unit{mHz}}{f_{\rm orb}}),
\end{equation}
where $\rho$ is the \snr, $T_{\rm obs}$ is the \lisa mission observation time and $f_{\rm orb}$ is the orbital frequency. The chirp mass
can be written as follows \citep[e.g.,][Eq.\ 25]{LEGWORK}
\begin{equation}
    \label{eq:chirp_mass}
    \mathcal{M}_c = \qty(\frac{5\pi}{48 n} \frac{c^5}{G^{5/3}} \dot{f}_n)^{3/5} (2 \pi f_{\rm orb})^{-11/5} \, F(e)^{-3/5},
\end{equation}
and therefore the chirp mass
uncertainty can be estimated as \citep[e.g.,][Eq.\ C3]{Wagg+2022}
\begin{equation}
    \label{eq:chirp_mass_uncertainty}
    \frac{\Delta \mathcal{M}_c}{\mathcal{M}_c} = \frac{11}{5} \frac{\Delta f_{\rm orb}}{f_{\rm orb}} + \frac{3}{5} \frac{\Delta \dot{f}_{\rm dom}}{\dot{f}_{\rm dom}} + \frac{3}{5} \frac{\Delta F(e)}{F(e)},
\end{equation}
where $f_{\rm dom}$ is the harmonic frequency with the strongest \snr ($f_{\rm dom} = n_{\rm dom} f_{\rm orb}$ and $n_{\rm dom} = 2$ for circular binaries) as this will provide the best measurement. We estimate the frequency derivative uncertainty, which is the dominant contribution to the chirp mass uncertainty, following \citet{Takahashi+2002},
\begin{equation}
    \frac{\Delta \dot{f}_{\rm dom}}{\dot{f}_{\rm dom}} = 6 \sqrt{5} \cdot \frac{1}{\rho} \left(\frac{1}{T_{\rm obs}} \right)^2 \frac{1}{\dot{f}_{\rm dom}},
    \label{eq:f_orb_dot_unc}
\end{equation}
and estimate the eccentricity uncertainty, $\Delta e$, based on the relative \snr of the two loudest harmonics \citep[e.g.,][]{Seto+2016:2016MNRAS.460L...1S,Lau+2020,Andrews+2020:2020ApJ...892L...9A}, such that
\begin{equation}
    \frac{\Delta e}{e} = \frac{1}{\rho_1} + \frac{1}{\rho_2},
\end{equation}
where $\rho_n$ are the \snr of the two loudest harmonics.

For the spatial distribution uncertainty, \citet{Takahashi+2002} provides an expression for the uncertainty on the distance, $D$, as follows
\begin{equation}
    \begin{split} \frac{\Delta D}{D} = 0.2 &\qty(\frac{10}{\rho}) \max\Bigg\{ 1, \qty(\frac{f_{\rm orb}}{\num{1.4e-3}\unit{Hz}})^{-11/3} \\ &\times \qty(\frac{\mc}{1 \msun})^{-5/3} \qty(\frac{T_{\rm obs}}{10 \unit{yr}})^{-2} \Bigg\},
    \end{split}
\end{equation}
and the uncertainty on the sky localisation, $\theta$, is given as follows \citep{Mandel+2018}
\begin{equation}
    \Delta \theta = 16.6\deg \, \qty(\frac{7}{\rho}) \qty(\frac{\num{5e-4}\unit{Hz}}{f_{\rm dom}}).
\end{equation}
We can write that the distance of a source from the Galactic plane as
\begin{equation}
    |z| = | D \sin b |,
\end{equation}
where $b$ is the Galactic latitude. Therefore, propagating the uncertainties and setting $\Delta b = \Delta \theta / \sqrt{2}$, the uncertainty on the distance from the plane can be written as
\begin{equation}\label{eq:z_unc}
    \Delta |z| = D \sqrt{\sin^2 b \qty(\frac{\Delta D}{D})^2 + \cos^2 b (\Delta b)^2}.
\end{equation}

\section{Supplemental tables}

Table~\ref{tab:detections_init} shows the number of detectable DCOs when using the \textit{formation} location of each DCO, while Table~\ref{tab:detections_final} shows the same number when using the \textit{present-day} location. These values are visualised in Figure~\ref{fig:n_detect}.

\begin{table*}
    \centering
    \input{tables/detections_init_pos}
    \caption{Number of detectable DCOs for a 8-year \lisa observation duration, an \snr limit of 12, and using the formation location of each DCO in \snr calculations. Each column shows the number for each DCO type. The values are the median from 2560 instances of the Milky Way and the uncertainties give the 5th and 95th percentiles. Each row shows the different model variations beyond the fiducial simulation, see Table~\ref{tab:models} and Section~\ref{sec:cosmic_explained} for explanations of each model.}
    \label{tab:detections_init}
\end{table*}

\begin{table*}
    \centering
    \input{tables/detections_final_pos}
    \caption{As Table~\ref{tab:detections_init}, but using the present-day positions of DCOs to calculate their \snr. Typically fewer systems are detectable in this case since supernova kicks often displace DCOs further from the Earth.}
    \label{tab:detections_final}
\end{table*}

\end{document}

%% file: tables/n_targets.tex
\begin{tabular}{lccccc}
\hline
Model & BHBH & BHNS & BHWD & NSNS & NSWD \\
 & [$10^5$] & [$10^5$] & [$10^4$] & [$10^4$] & [$10^5$] \\
\midrule
Fiducial & 4.55 & 2.17 & 1.60 & 2.34 & 6.28 \\
C14 $q_{\rm crit}$ & 4.86 & 2.67 & 0.45 & 2.59 & 5.26 \\
Optimistic CE & 11.56 & 5.38 & 15.39 & 17.73 & 9.83 \\
$\alpha_{\rm CE} = 0.1$ & 4.55 & 1.53 & 0.29 & 1.40 & 0.13 \\
$\alpha_{\rm CE} = 10$ & 0.69 & 1.46 & 0.008 & 15.83 & 2.20 \\
M25 remnants & 0.46 & 1.81 & 0.005 & 0.73 & 6.08 \\
No ECSN & 4.54 & 2.18 & 1.59 & 2.32 & 13.98 \\
\hline
\end{tabular}

%% file: tables/detections_init_pos.tex
\begin{tabular}{lccccc}
\toprule
Model & BHBH & BHNS & BHWD & NSNS & NSWD \\
\midrule
Fiducial & $13.1^{+9.8}_{-6.2}$ & $8.0^{+6.5}_{-3.7}$ & $1.3^{+0.5}_{-0.4}$ & $0.8^{+0.7}_{-0.3}$ & $110^{+25}_{-22}$ \\
Claeys+2014 critical mass ratios & $13.4^{+10.3}_{-6.2}$ & $9.6^{+7.3}_{-4.3}$ & $0.4^{+0.2}_{-0.1}$ & $0.9^{+0.6}_{-0.4}$ & $72^{+20}_{-17}$ \\
Optimistic CE & $39^{+27}_{-17}$ & $19.1^{+16.2}_{-8.1}$ & $9.6^{+4.4}_{-3.1}$ & $6.3^{+5.5}_{-2.7}$ & $173^{+50}_{-38}$ \\
Low efficiency CE, $\alpha_{\rm CE} = 0.1$ & $13.8^{+12.8}_{-6.5}$ & $4.8^{+6.2}_{-2.6}$ & $0.3^{+0.2}_{-0.1}$ & $0.7^{+0.5}_{-0.3}$ & $1.0^{+0.7}_{-0.5}$ \\
High efficiency CE, $\alpha_{\rm CE} = 10$ & $2.1^{+1.9}_{-1.0}$ & $5.8^{+3.8}_{-2.5}$ & $0.0^{+0.0}_{-0.0}$ & $6.9^{+3.7}_{-2.6}$ & $24.4^{+12.0}_{-8.8}$ \\
Maltsev+2025 remnant masses & $0.6^{+0.9}_{-0.4}$ & $7.3^{+4.8}_{-3.0}$ & $0.0^{+0.0}_{-0.0}$ & $0.3^{+0.3}_{-0.1}$ & $117^{+27}_{-22}$ \\
No electron-capture supernovae & $13.0^{+9.9}_{-6.1}$ & $8.0^{+6.2}_{-3.7}$ & $1.3^{+0.6}_{-0.4}$ & $0.7^{+0.5}_{-0.3}$ & $158^{+67}_{-49}$ \\
\bottomrule
\end{tabular}

%% file: tables/detections_final_pos.tex
\begin{tabular}{lccccc}
\toprule
Model & BHBH & BHNS & BHWD & NSNS & NSWD \\
\midrule
Fiducial & $11.6^{+9.1}_{-5.9}$ & $5.5^{+5.6}_{-2.8}$ & $1.2^{+0.6}_{-0.4}$ & $0.5^{+0.6}_{-0.3}$ & $99^{+25}_{-20}$ \\
Claeys+2014 critical mass ratios & $12.0^{+9.7}_{-6.1}$ & $6.8^{+5.9}_{-3.6}$ & $0.4^{+0.2}_{-0.1}$ & $0.6^{+0.5}_{-0.3}$ & $63^{+19}_{-15}$ \\
Optimistic CE & $35^{+25}_{-16}$ & $13.9^{+13.0}_{-7.0}$ & $9.4^{+4.3}_{-3.1}$ & $5.1^{+5.0}_{-2.4}$ & $161^{+49}_{-37}$ \\
Low efficiency CE, $\alpha_{\rm CE} = 0.1$ & $12.6^{+11.7}_{-6.1}$ & $3.7^{+5.5}_{-2.2}$ & $0.3^{+0.2}_{-0.1}$ & $0.2^{+0.2}_{-0.1}$ & $1.0^{+0.6}_{-0.5}$ \\
High efficiency CE, $\alpha_{\rm CE} = 10$ & $2.0^{+1.8}_{-1.0}$ & $4.8^{+3.3}_{-2.2}$ & $0.0^{+0.0}_{-0.0}$ & $5.9^{+3.2}_{-2.4}$ & $22.1^{+11.4}_{-8.2}$ \\
Maltsev+2025 remnant masses & $0.6^{+0.9}_{-0.4}$ & $4.1^{+3.7}_{-2.2}$ & $0.0^{+0.0}_{-0.0}$ & $0.3^{+0.3}_{-0.1}$ & $104^{+26}_{-21}$ \\
No electron-capture supernovae & $11.7^{+9.2}_{-5.7}$ & $5.6^{+5.3}_{-2.9}$ & $1.3^{+0.6}_{-0.4}$ & $0.3^{+0.4}_{-0.2}$ & $106^{+58}_{-39}$ \\
\bottomrule
\end{tabular}